\documentclass[10pt, aip, twocolumn, nofootinbib, amssymb, amsmath]{revtex4-1}
\usepackage[paperwidth=210mm,paperheight=297mm,centering,hmargin=1.2cm,vmargin=1.7cm]{geometry}
\usepackage{gensymb}
\usepackage[caption=false]{subfig}
\usepackage{graphicx}
\usepackage[dvipsnames]{xcolor}
\usepackage{hyperref}
\usepackage{listings}

\graphicspath{{figures/}}
\begin{document}

%\title{\large Examining the \textit{Rule of Four} Anomaly in Inorganic Databases}

\title{\large The \textit{rule of four}: anomalous stoichiometries of inorganic compounds}

\author{Elena Gazzarrini}
\affiliation{Theory and Simulation of Materials (THEOS) and National Center for Computational Design and Discovery of Novel Materials (MARVEL), École Polytechnique Fédérale de Lausanne, CH-1015 Lausanne, Switzerland}

\author{Rose K. Cersonsky}
\affiliation{Department of Chemical and Biological Engineering, University of Wisconsin - Madison, Madison, Wisconsin, USA}

\author{Marnik Bercx}
\affiliation{Theory and Simulation of Materials (THEOS) and National Center for Computational Design and Discovery of Novel Materials (MARVEL), École Polytechnique Fédérale de Lausanne, CH-1015 Lausanne, Switzerland}

\author{Carl S. Adorf}
\affiliation{Theory and Simulation of Materials (THEOS) and National Center for Computational Design and Discovery of Novel Materials (MARVEL), École Polytechnique Fédérale de Lausanne, CH-1015 Lausanne, Switzerland}

\author{Nicola Marzari}
\affiliation{Theory and Simulation of Materials (THEOS) and National Center for Computational Design and Discovery of Novel Materials (MARVEL), École Polytechnique Fédérale de Lausanne, CH-1015 Lausanne, Switzerland}

\date{26 July 2023}

\begin{abstract}

Why are materials with specific characteristics more abundant than others?
This is a fundamental question in materials science and one that is traditionally difficult to tackle, given the vastness of compositional and configurational space.
We highlight here the anomalous abundance of  inorganic compounds whose primitive unit cell contains a number of atoms that is a multiple of four.
This occurrence --- named here the \textit{rule of four} --- has to our knowledge not previously been reported or studied.
Here, we first highlight the rule's existence, especially notable when restricting oneself to experimentally known compounds, and explore its possible relationship with established descriptors of  crystal structures, from symmetries to energies.
We then investigate this relative abundance by looking at structural descriptors, both of global (packing configurations) and local (the smooth overlap of atomic positions) nature.
Contrary to intuition, the overabundance does not correlate with low-energy or high-symmetry structures; in fact, structures which obey the \textit{rule of four} are characterized by low symmetries and loosely packed arrangements maximizing the free volume. 
We are able to correlate this abundance with local structural symmetries, and visualize the results using a hybrid supervised-unsupervised machine learning method.
%This result points to the phenomenon being symmetric in nature, rather than a signature of compounds exhibiting ideal target properties, and has therefore limited implications on materials design.
\end{abstract}

\maketitle

\section{Introduction}

Computational materials discovery is a fast-growing discipline leading to innovation in many fields.
Within a specific technological sector (i.e., communications, renewable energies, medical), the choice of material is critical for the long-lasting success of the given product.
Therefore, it is important -- and of fundamental interest -- to efficiently identify materials' structural and energetic characteristics through materials' data analysis to select structures for innovative applications.
The emerging field of materials informatics has demonstrated its potential as a springboard for materials development, alongside first-principles techniques such as density-functional theory (DFT) \cite{PhysRev.136.B864, dft_marzari}.
The increase in computational power, together with large-scale experimental \cite{high_throughput_exp} and computational high-throughput studies \cite{high_throughput_theo}, is paving the way for data-intensive, systematic approaches to classify materials' features and to screen for optimal experimental candidates.
In addition, the collection of statistical methods offered by machine learning (ML) has accelerated these efforts, both within fundamental and applied research \cite{Vasudevan2019, Pilania2013, Rupp2012, Sanchez2018, bartok_machine_2017, de_comparing_2016}.

\begin{figure}[h!]  
  \centering
  \includegraphics[width=0.7\columnwidth]{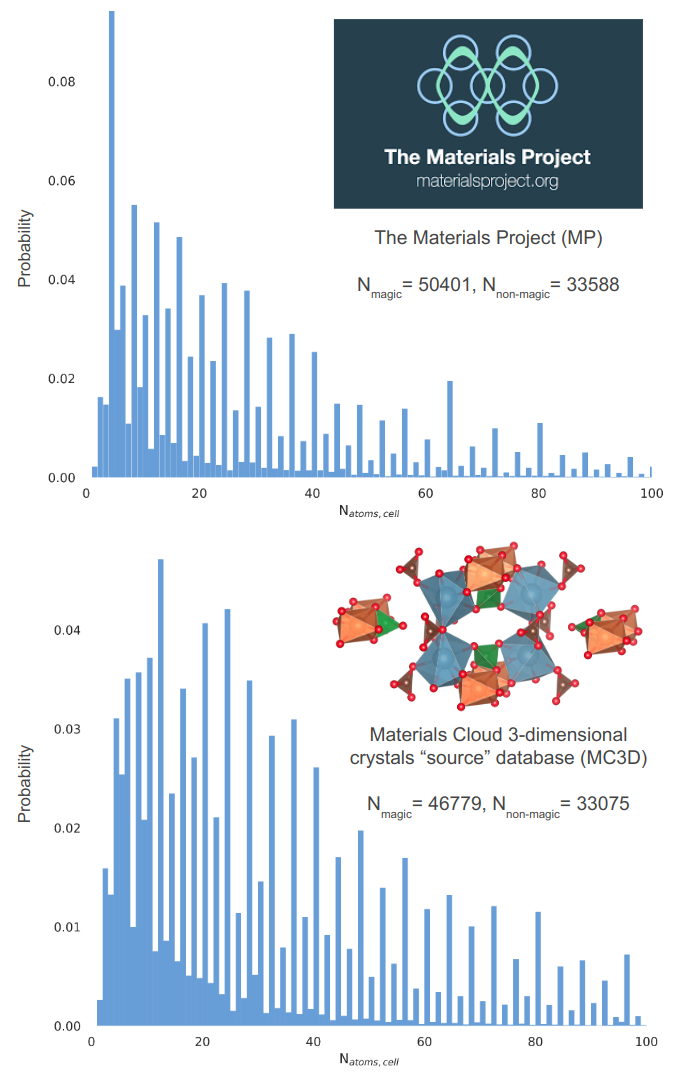}
  \caption{
      The \textit{rule of four}. The two datasets (the Materials Project (MP) \cite{jain_commentary_2013} and the Materials Cloud 3-dimensional crystal structures `source' database (MC3D-source))~\cite{mc3d}) contain a disproportionate amount of compounds with a primitive unit cell containing multiples of 4 atoms. 
  }
  \label{fig:natom_occ}
\end{figure}

However, the success of these endeavors is ultimately limited by the quality and diversity of the data serving as the underlying data source.
Understanding the space of materials spanned by a dataset is integral to data-driven materials searches or machine-learning workflows.
Thus, when anomalous correlations arise in datasets, it is useful to understand and investigate the origins, and potential implications, of such peculiarities.
We use here the name \textit{rule of four} (RoF) to describe the unusually  high relative abundance of structures with primitive unit cells containing a multiple of 4 atoms. This occurrence is explored within two different databases of inorganic crystal structures: the Materials Project (MP) \cite{jain_commentary_2013}database, which contains crystal structures that have been relaxed with first-principles calculations starting from experimental databases or from structure-prediction methods, and the Materials Cloud 3-dimensional crystal structures `source' database (MC3D-source); this latter combines experimental structures from the crystallographic open database\,(COD)\,\cite{cod1, cod2, cod3, cod4}, the inorganic crystal structures database\,(ICSD)\,\cite{icsd_n} and the materials platform for data science\,(MPDS) \footnote{Note that for the ICSD and COD, occasionally some theoretically predicted structures can also be present, see section~I in the supplementary information for more details.}. 
 Figure~\ref{fig:natom_occ} is a visual representation of this striking abundance, while Table~\ref{table:rof_quantify} demonstrates the RoF by comparing the relative abundance of structures with primitive unit cells made up of multiple of 3, 4, 5, 6 and 7 atoms. 
%\, which contains the Pauling File \cite{Pauling} database (we are not using the 3DCD's set of dft-relaxed structures). 

% \begin{table}[htbp!] 
% \centering
% \begin{tabular} { | c | c | c | c | c | c | c |}
% \hline
%  & \textit{m-1} &\textbf{\textit{m=4}} & \textit{m+1} & \textit{m+2} & \textit{m+3} \\
% \hline
% \hline
% \textbf{MC3D-source} & 36.57  & \textbf{58.58} & 20.89 & 30.99 & 12.51\\
% \hline
% \textbf{Materials Project}  & 32.38 & \textbf{60.01} & 18.41 & 26.82 &  12.43\\
% \hline

% \end{tabular}
% \vspace{0.2 cm}
% \caption{ Absolute percentages of unit cells in both databases that contain a number of atoms multiple to the column header (i.e. the first columns indicates the overall percentage of structures containing 3 or multiple of 3 atoms in their primitive unit cell). 
% }
% \label{table:rof_quantify}
% \end{table}

\begin{table}[htbp!] 
\centering
\begin{tabular} { | c | c | c | c | c | c | c |}
\hline
multiple of & 3 & \textbf{4} & 5 & 6  & 7 \\
\hline
\hline
\textbf{Materials Project}  & 32.38 & \textbf{60.01} & 18.41 & 26.82 &  12.43\\
\hline
\textbf{MC3D-source} & 36.57  & \textbf{58.58} & 20.89 & 30.99 & 12.51\\
\hline

\end{tabular}
\vspace{0.2 cm}
\caption{
Percentages of structures in the MP and MC3D-source databases whose primitive unit cells contain a number of atoms that is a multiple of the column header. The RoF emerges from the higher abundance of structures with a primitive unit cell containing a multiple of 4 atoms. Primitive unit cells with a number of atoms that is a multiple of two or more headers will contribute to each column; hence, the percentages will sum to more than 100.
}
\label{table:rof_quantify}
\end{table}

Within the context of this study, we will label a structure that belongs to the subset of structures with a unit cell size multiple of four as a \emph{magic} structure, and one that does not belong to the subset as a \emph{non-magic} structure. 
In Figure~\ref{fig:natom_occ} the $x$ axis is capped at 100 atoms to best represent the RoF, as respectively  97.51\% and 91.00\% of structures in the MP and in the MC3D-source databases contain 100 atoms or less (the largest cell in the MC3D-source database contains 4986 atoms).

\begin{figure}
    \centering
    \includegraphics[width=\linewidth]{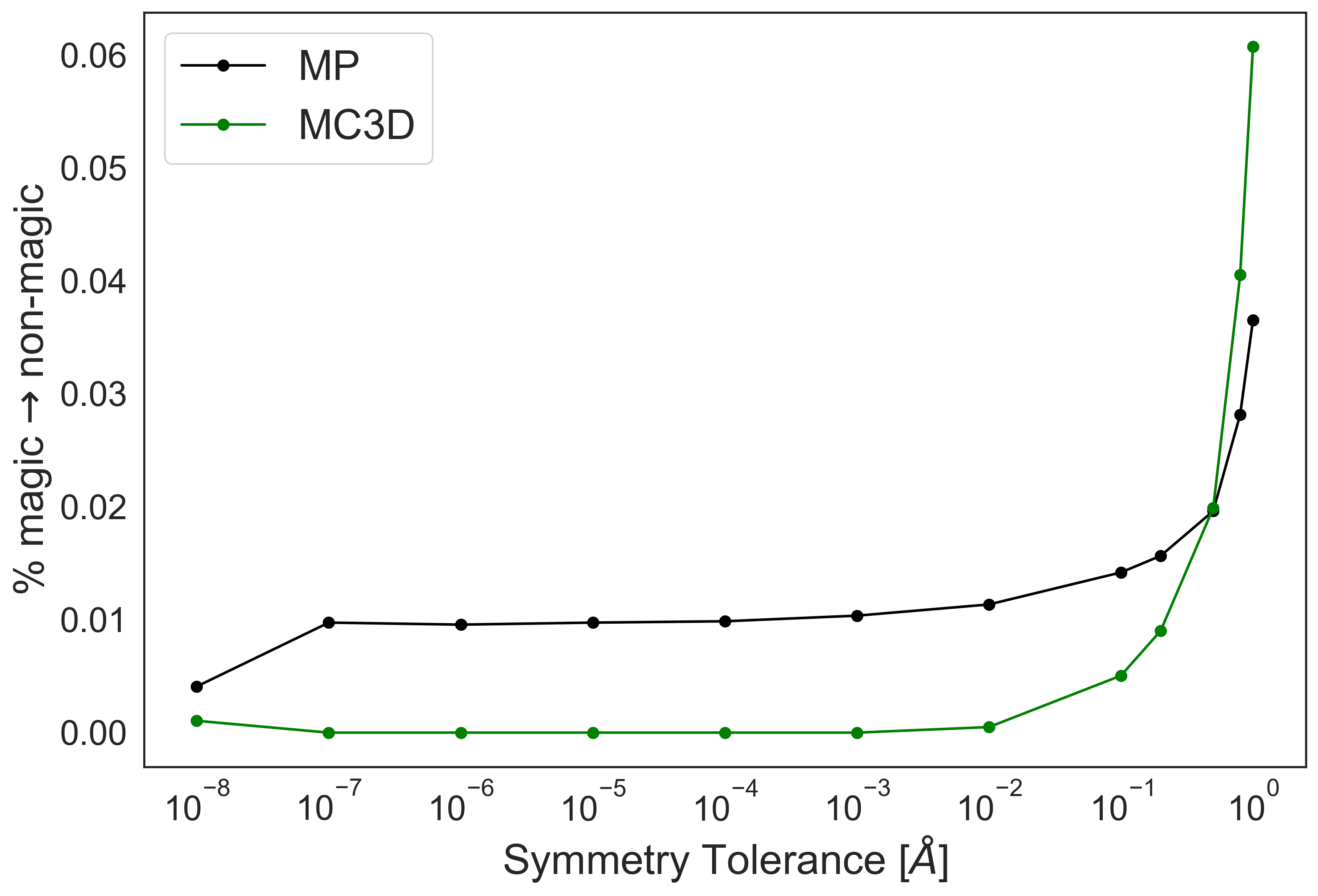}
    \caption{Percentage of \emph{magic} structures that become labelled \emph{non-magic} as a function of the symmetry tolerance parameter used for reduction to the primitive cell.
    The black and green lines correspond to structures in the MP and MC3D-source datasets, respectively.
    At typical symmetrization parameters, there is little to no change in the number of \emph{magic} structures (roughly 1\% of \emph{magic} structures go to \emph{non-magic}).
    At larger symmetrization parameters ($\approx$ 1\AA), this increases to roughly 6\% based upon the large deviations allowed in considering sites as symmetrically equivalent.}
    \label{fig:flip}
\end{figure}

Before delving into a more extensive analysis, we want to rule out that the RoF is simply an artifact of how structures are mathematically described, or of how this description is curated and processed for storage in the aforementioned databases.
When materials structure datasets are prepared, it is standard procedure to `primitivise' unit cells, i.e., to reduce the unit cell to its minimum volume. As many conventional unit cells contain exactly four times the number of atoms that would be found in their respective primitive unit cell, it could be expected that misclassifying conventional unit cells as primitive ones could lead to an artificial emergence of the RoF. Both the MP and MC3D-source databases obtain the primitive unit cell using the \texttt{spglib} software~\cite{spglib}. When primitivizing the structure, one needs to set the \texttt{symprec} tolerance parameter, which allows for slight deviations in the atomic positions stemming from thermal motion or experimental noise.
% In either case, one can choose two tolerance parameters: \texttt{symprec} and \texttt{angle\_tolerance}, that allow for slight deviations in the atomic positions stemming from thermal motion or experimental noise.
% For \texttt{angle\_tolerance}, we found that doubling the already high default value of this parameter ($5\degree$) had little to no effect on the number of \textit{magic} structures (SEE SI-ADD REF).
To rule out that the primitivization is the source of the emergence of the RoF, we show in Fig.~\ref{fig:flip} that changing the \texttt{symprec} (1E-8 to 1E-1\AA) parameter has little effect on the RoF distribution, converting around 1\% of \emph{magic} structures into \emph{non-magic} ones.
It is only when one increases the \texttt{symprec} to unreasonably large values (close to 1\AA) that the slope changes -- this is expected, as using such a large tolerance effectively considers sites with the same element that should be different as identical, producing primitive unit cells with a reduced number of sites, but which no longer correctly describe the structure.

Encouraged by these results, we decide to probe the RoF more deeply and attempt to understand its origins and impact. First, we examine the RoF with respect to traditional materials science metrics, including energies and symmetries, and uncover that the RoF is largely correlated with loosely-packed polyatomic systems.
We then use symmetry-adapted machine learning techniques to relate the RoF to local atomic environments and determine that it has only little implications for energetic stability.
We then manage to correctly classify the RoF by only providing the algorithm with information on local structural symmetry rather than a global one. 

% The investigation carried out is composed of three main steps: (1) demonstrating that the RoF is not an artifact of how the unit cells of the structures are defined in the database, (2) determining the effects of the RoF on the compounds' physical properties, and, conversely, (3) understanding any correlated structural characteristics. 

% We find that the abundance of \emph{magic} structures within the datasets is able to be predicted from local features and symmetries only, and is correlated with low-symmetry configurations and lower packing fractions, rather than with stable energetic configurations.

\section{Results and discussion}

Within this study, we make sure that the data is sufficiently diverse for the training set to cover the whole design space \cite{Ajiboye2015} by procuring the structural data from open and FAIR repositories\cite{fair1, schmidt_recent_2019,fair2}; the same analytical workflow is applied to two different databases of bulk, crystalline, stoichiometric compounds.
One database is the Materials Project, which contained 83\,989 data entries obtained via high-throughput DFT calculations as of 10/18/2018, corresponding to the \texttt{mp\_all\_20181018} dataset retrieved with the  \texttt{matminer.datasets} module~\cite{data_access}. The other data source, the MC3D-source, contains 79\,854 unique structures extracted from the MPDS, ICSD and COD, which have been curated via an AiiDA \cite{aiida} workflow, as explained in Section~I of the SI. 

\subsection{Energetic stability}
\label{sec:energy}

% As databases, to a certain extent, contain more structures that are locally stable (\textit{e.g.}, with lower energy than other similar conformations)
We first test whether the RoF is correlated with energetic stability, as this would provide a straightforward explanation for the phenomenon.
To test this assumption, we analyze the information contained in the MP dataset, namely the formation energy per atom within each compound. This is the energy of the compound with respect to standard states (elements), normalized per atom \footnote{For example, for Fe2O3 the formation energy is [E(Fe2O3) - 2E(Fe) - 3/2E(O2)]/5}. 
It is computed at a temperature of 0 K and a pressure of 0 atm. This quantity is often a good approximation for formation enthalpy at ambient conditions, where a negative formation energy implies stability with respect to elemental compounds.

\begin{figure}[ht] 
    \centering
    \includegraphics[width=1\columnwidth]{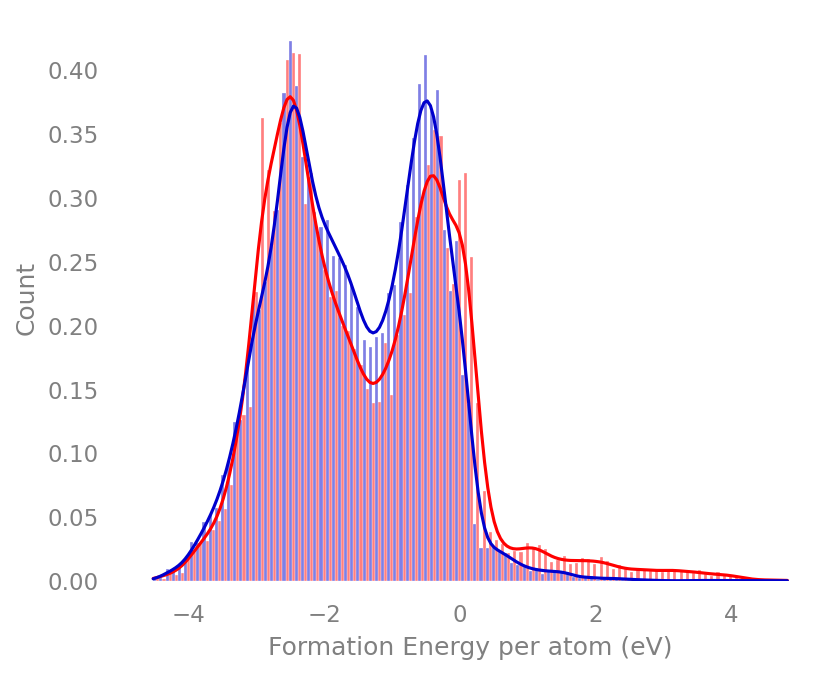}
\caption{
    Probability distribution of formation energies for the 83\,989 compounds from the Materials Project, normalized for each subgroup.
    \emph{Magic} compounds are colored in red and \emph{non-magic} are colored in blue.
}
\label{fig:E_discrete}
\end{figure}

Our initial results provide no evidence of a correlation between \emph{magic} compounds and their energetic stability, as shown in Figure \ref{fig:E_discrete}.
Nevertheless, it does appear that structures obeying the RoF have a longer positive tail of large formation energies, seen towards the bottom right of the figure.

% \begin{figure*}[ht] 
%     \centering
%     \includegraphics[width=.98\textwidth]{energies/E_pcovr.png}  
% \caption{
%  PCovR representation of the MP dataset with a mixing parameter of $\beta$=0.5.
% The model is regressed on the formation energy per atom. The three plots contain the same data, [a] represented through a kernel density probability distribution (the \emph{magic} subset is coloured in red and the \emph{non-magic} one in blue), [b] coloured according to the subset classification and [c] according to the formation energy per atom. }
% \label{fig:pcovr_E}
% \end{figure*}

However, this result can be misleading -- it does not take into consideration the large variance in structural composition across the database -- and we must aim to compare the energies of similar structures within the \emph{magic} and \emph{non-magic} subsets, as we will do in later sections.

\subsection{Correlation with symmetry descriptors}
\label{sec:symm}

\begin{figure*}[ht]
    \centering
    \subfloat[MC3D-source]{
        \includegraphics[width=0.48\textwidth]{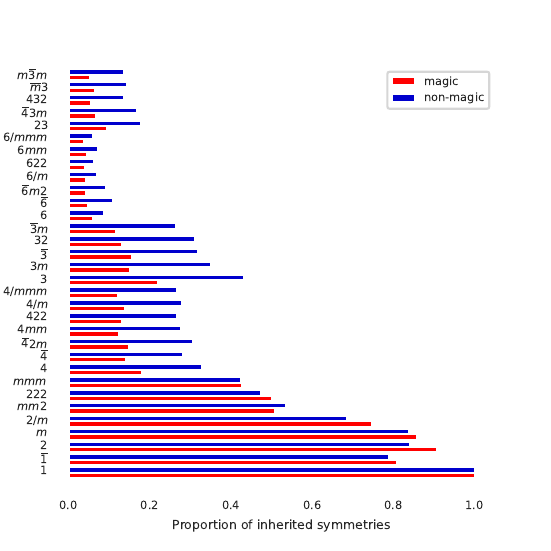}
    }
    \subfloat[MP]{
        \includegraphics[width=0.48\textwidth]{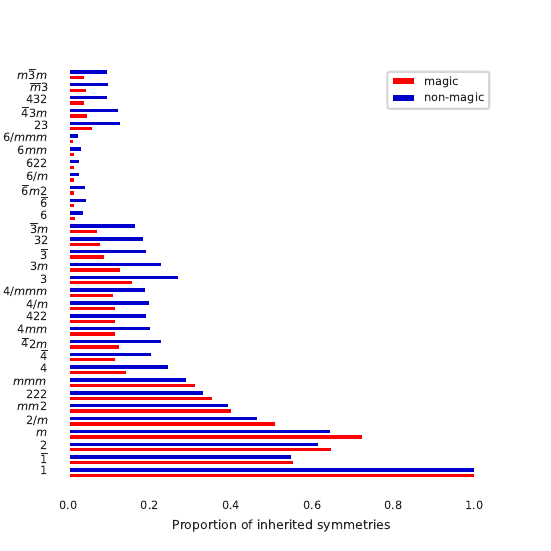}
    }
  \caption{
      Proportion of structures in both databases ([a]  MC3D-source and [b] MP) that belong to each point group represented on the $y$ axis, counted based on their inherited symmetries. 
      \emph{Magic} compounds are coloured in red, while \emph{non-magic} ones in blue.
  }
  \label{fig:inh_symm}
\end{figure*}

The crystal symmetries of compounds -- defined by the set of symmetry operations that, when performed, leave the structure unchanged -- are captured in crystals by their space groups and point groups.
Higher symmetry space groups inherit the symmetry operations of their `parent' point groups; for example, cubic space groups inherit the one-fold, two-fold, and four-fold rotational symmetries \footnote{For the interested reader, the concept of inherited symmetry is enumerated nicely in Fig.\,5 of the book chapter by Hestenes \cite{dorst_point_2002}.}. 
Figure~\ref{fig:inh_symm} shows histograms of inherited symmetries and their relative abundance within each of the two sets (\emph{magic} in red and \emph{non-magic} in blue).
The point groups are ordered from the ones with the least number of symmetry operations (bottom) to the highest order ones (top).
Symmetry groups that are equally represented in both sets (i.e. 1-rotation, since all compounds are invariant to the simplest symmetry) have tails of equal length, whereas symmetries seen in a larger percentage of \emph{magic} structures have a red tail to the right of the histogram.
% All compounds are invariant to the 1-rotation symmetry, represented by space group `1'; this is the reason why the relative abundance for this particular subgroup is 100\%.
From Figure~\ref{fig:inh_symm}, the relative abundance of \emph{non-magic} structures in the high symmetry point groups emerges, while on the contrary most \emph{magic} structures in both databases are grouped in the lowest symmetry point groups ($2$, $m$, $2/m$, $mm2$, $222$ and $mmm$), which generally contain a relative abundance of them apart from one exception (the MC3D-source presents a slightly higher relative abundance of \emph{non-magic} structures in the $mm2$ point group). 
This analysis shows how  4-fold symmetry is \emph{not} a determining descriptor to classify the phenomenon. 

% , nor FCC and HCP symmetries strongly characterize \emph{magic} structures. 

\begin{figure*}[ht] 
\centering
    \subfloat[MC3D-source]{
        \includegraphics[width=.95\textwidth]{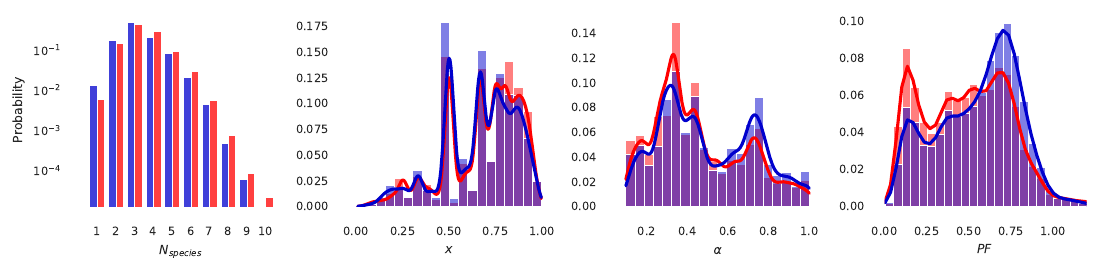}
    }\\
    \subfloat[MP]{
        \includegraphics[width=0.95\textwidth]{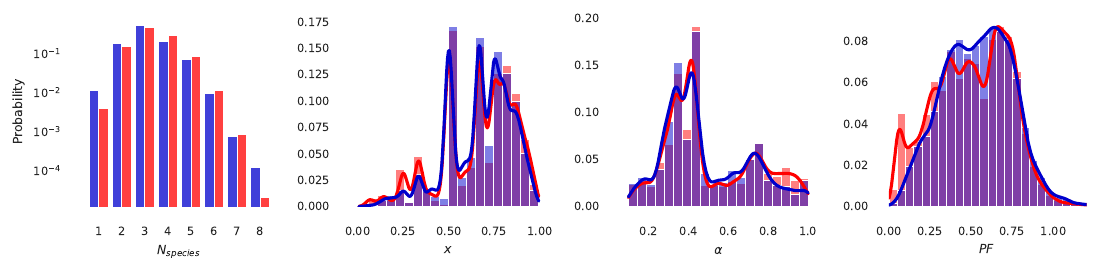}
    }
\caption{
    Different geometric properties of each compound are analysed for the (a) MC3D-source and (b) MP databases.
    From left to right, the plots represent the distribution of the number of elemental species ($N_{species}$), the relative abundance of small (${N_{S}}$) to large ($N_{L}$) radii ($x$), the ratio between smallest (${R_{S}}$) and largest (${R_{L}}$) atomic radii ($\alpha$) and the packing fraction ($PF$) for compounds with a unit cell size between 0 and 100 atoms.
    All of the results are plotted for the two sets, \emph{magic} (red) and \emph{non-magic} (blue), with the probability normalized to each set.} \label{fig:packing}
\end{figure*}

The symmetrical `disorder' -- or higher asymmetry -- characterising \emph{magic} compounds may be caused by a more heterogeneous composition of atoms, as compared to \emph{non-magic} compounds.
We can quantify this heterogeneity by counting the number of atomic species ( $N_{species}$) (first column of Figure \ref{fig:packing}) composing the structures: from this analysis we see that \emph{magic} materials are mostly composed of 4 or more elements, while \emph{non-magic} structures present a larger abundance of simpler composition, containing more often 1, 2, or 3 elements.
Another property that emerges from our analysis and is more evident in the MP dataset (second column of Figure \ref{fig:packing}(b)) is the relative scarcity of smaller atomic radii within \emph{magic} compounds, as often defined by the parameter $x = \frac{N_{S}}{N_{S} + N_{L}}$, where $N_{S}$ and $N_{L}$ are the counts, in a given structure, of the smallest and largest radii respectively.
The scarcity of small radii in \emph{magic} compounds (lower $x$ parameter) partly explains the lower symmetries that characterise them, as no atoms will easily be inserted as `interstitial' elements in a given structure. On the other side, the MC3D-source dataset also presents a peak in higher values for $x$, in which case the largest atoms are much less than the smallest ones.
In this case, the smallest atoms might be seen as `imperfections', lowering the overall structural symmetry of point groups analyzed in Figure \ref{fig:inh_symm}. 

% On this account, we should instead assess the prevailing structural symmetry considering only the largest atoms.
% The reduction procedure is illustrated in Figure~\ref{fig:ovito}, produced with the OVITO software \cite{ovito}, to test whether \emph{magic} reduced structures exhibit symmetries consistent with close packing structures, i.e. Face Centered Cubic (FCC) or Hexagonal Close Packed (HCP). 

% \begin{figure}[htbp!]
%     \centering
%     \includegraphics[width=0.5\columnwidth]{discrete/ovito.png}
%   \caption{
%       Procedure for reducing the crystal configuration to the largest atoms.
%       For example, Cu$_{4}$Cl$_{4}$O$_{4}$ (top left) is reduced to  Cu$_{4}$ (top right), since Cu has the largest atomic radius, while Cl$_{24}$O$_{30}$S$_{4}$Sb$_{8}$ (bottom left) is reduced to Sb$_{8}$ (bottom right).
%   }
%   \label{fig:ovito}
% \end{figure}

% Table I of the SI shows that, even in the case of reduced structures, the percentage of closely packed-resembling \emph{magic}  structures never exceeds 50\%.
% % Therefore, \emph{magic} structures are not strongly characterised by either FCC or HCP symmetries.
% Furthermore, the low-symmetry profile of \emph{magic} structures is confirmed by the fact that \emph{magic} compounds are mostly monoclinic and orthorhombic, and they  appear more prevalent in primitive lattice centerings. This results after a Bravais lattices analysis is conducted, which can be found in Section IV of the SI. 
The low abundance of smaller radii in \emph{magic} compounds would likely lower the overall crystal symmetry of \emph{magic} compounds.
In general, the symmetry type of atomic crystal systems is strictly linked to packing mechanisms \cite{PhysRevLett.107.155501, Torquato_2009}. 
While the mathematical problem of sphere packing is not hard to pose (Kepler conjecture), it was historically difficult to prove \cite{Hales}, and the complexity of its solution rises exponentially with polydispersity \cite{Torquato_2018}. 
% Already binary \cite{Hopkins_2011} and ternary \cite{koshoji2021densest} mixtures increase the complexity of the dense packing problem significantly.
% It is therefore desirable to devise a set of intensive parameters that can characterize packings well.
Despite this, a qualitative analysis of \emph{magic} configurations shows that they contain chemical elements whose size variance is much greater compared to the variance in the \emph{non-magic} population.

This size variance is quantified by the parameter $\alpha = \frac{R_{S}}{R_{L}}$ (where $R_{L}$ is the radius of the largest radius and $R_{S}$ of the smallest one), namely the ratio between the smallest and the biggest atomic radii within each compound (third column of Figure \ref{fig:packing}).
% It can be noticed how the MP dataset presents an abundance of \emph{magic} structures where the smallest to largest ratio approaches the unity; this feature might characterise \emph{magic} mono-atomic compounds.
It can be noticed how the MP dataset presents an abundance of \emph{magic} structures with the smallest to largest ratio between 0.8 and 1; this feature characterises the lower variance in elements that make up \emph{magic} compounds.
\emph{Magic} compounds from the MC3D-source exhibit a greater standard deviation between largest and smallest atoms, with the $\alpha$ parameter presenting a peak at around 0.35; this finding suggests the presence of very small radii filling the interstitial spaces, which contribute to keeping the symmetry of \emph{magic} compounds low. 
The packing fraction (PF), defined as $\textrm{PF} = \frac{\textrm{V}_{tot, atoms}}{\textrm{V}_{cell}}$, is another related property of sphere packing. 
This quantity is noticeably lower (with peaks at values around 0.1 - 0.2) for \emph{magic} structures, as can be seen in the last column of Figures~\ref{fig:packing} (a) and (b), pointing away from packing arguments as the cause of this database anomaly \cite{PhysRevLett.107.155501, Torquato_2009}.
The sharp red peaks in $PF$ might characterise disordered compounds such as metal-organic frameworks (MOFs) and other porous materials, which have been determined to be outliers for the MC3D-source dataset.  
This aligns with the thesis of Hopkins \cite{Hopkins_2011}, namely that entropic (free-volume maximizing) particle interactions contribute to the structural diversity of mechanically stable and ground-state structures of atomic, molecular, and granular solids.

\subsection{Employing symmetry-adapted descriptors for further insight}
\label{subsec:soap}
Up until this point we have employed classical techniques for analyzing crystal structures; here, we aim to understand the RoF using modern data-driven techniques.
In the field of atomistic modeling, it has been common, albeit non-trivial, to represent crystal structures through symmetrized density correlations \cite{bartok_machine_2017, bartok_representing_2012,behler_generalized_2007} in order to predict broad swaths of materials properties.
Here, we represent the compounds using the Smooth Overlap of Atomic Positions (SOAP) \cite{bartok_representing_2012}, a popular ML representation for structure-energy relations that contains information on the average three-body local environment for atomic arrangements.
SOAP vectors provide an avenue for a statistical analysis on local environments, offering a robust framework through which we can explore and visualize the chemical and configuration space of the materials studied \cite{musil_efficient_2021}.
We use two parameterizations of SOAP vectors, detailed in Section II of the SI: one that uses separate channels to represent different chemical species and another that ignores the chemical identities in order to highlight the geometry of the local symmetry.
The former, from hereon called the \emph{species-tagged} representation, is necessary in energetic analysis, as similar geometry symmetries can correspond to wildly different energetics given the elements present; however, this representation is computationally cumbersome (roughly 100\,000 sparse features for each compound, from which we take a diverse subset of 2\,000 features).
Thus, in later analyses where the chemical identities play a smaller role, it is beneficial and conceptually more straightforward to use the more lightweight, latter representation (roughly 80 features for each compound), hereon called the \emph{species-invariant} representation.

\begin{figure}[ht]
\centering
\includegraphics[width=.7\columnwidth]{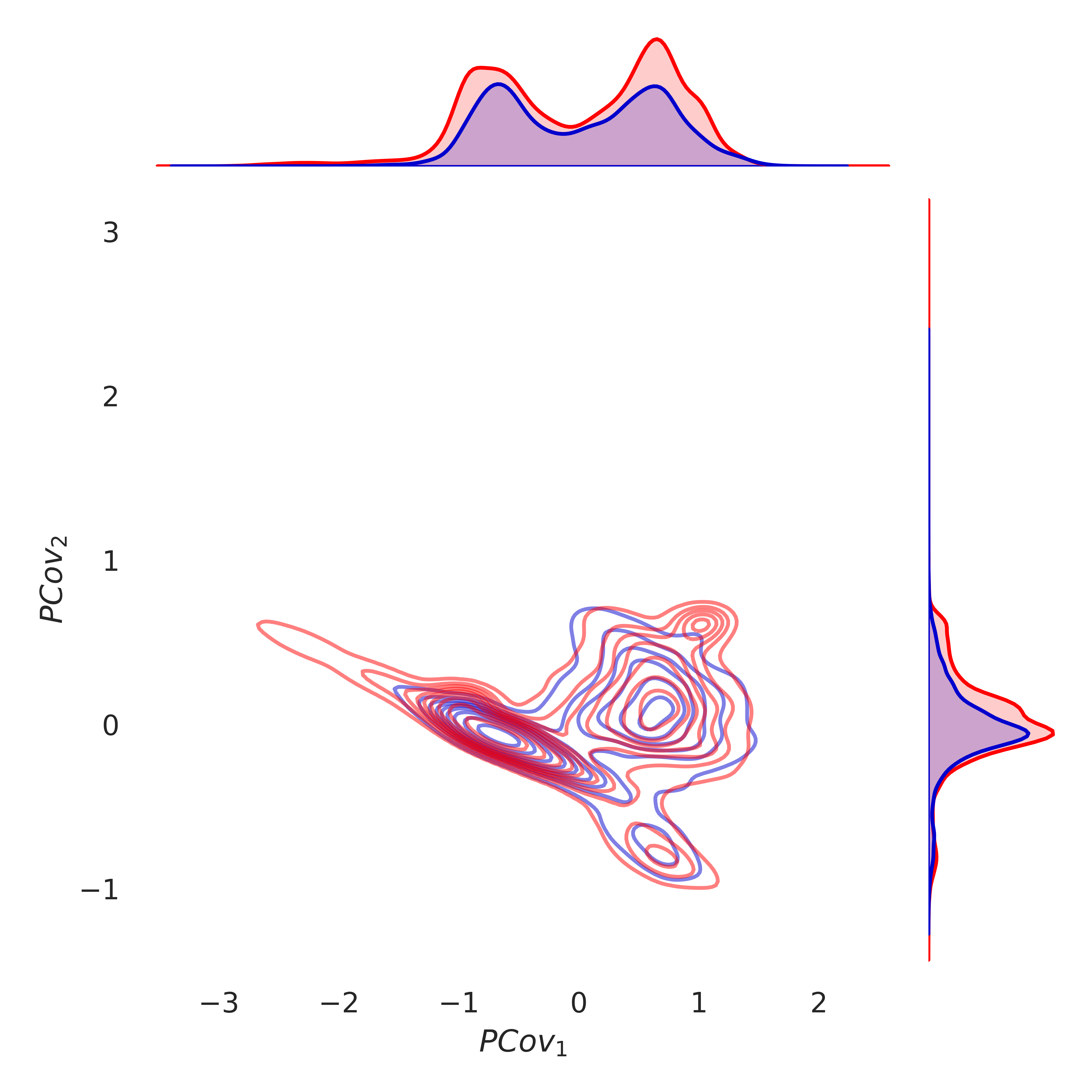} % Just stack two includegraphics!
\includegraphics[width=.9\columnwidth]{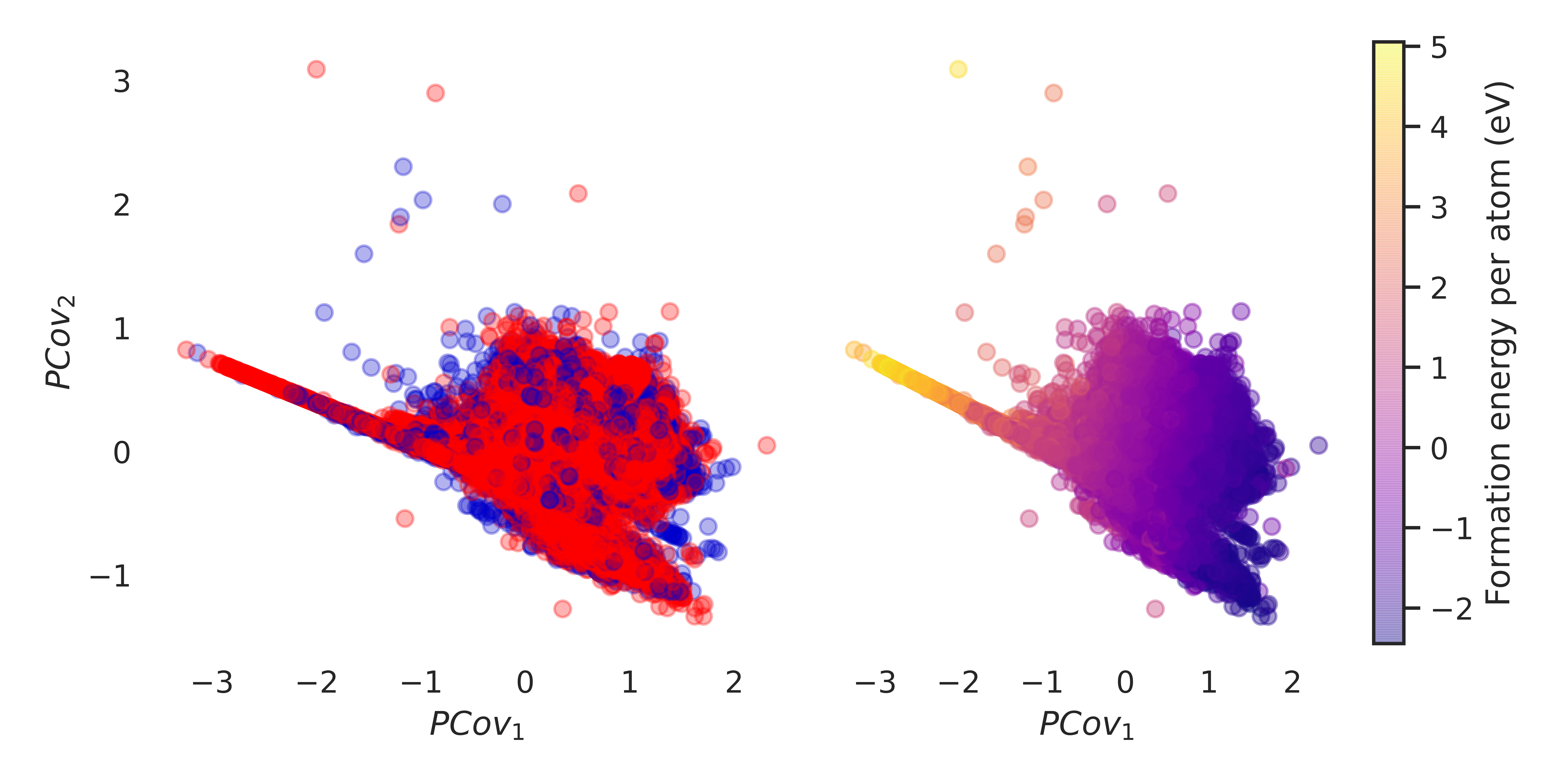}
\caption{PCovR representation of the MP dataset with a mixing parameter of $\beta$=0.5.
The model is regressed on the formation energy per atom.
The three plots contain the same data, represented on the top through a kernel density probability distribution (the \emph{magic} subset is coloured in red and the \emph{non-magic} one in blue), coloured according to the subset classification (lower left) and according to the formation energy per atom (lower right).}
\label{fig:pcovr_E}
\end{figure}

Earlier, we noted that simply presenting a histogram of \emph{magic} and \emph{non-magic} energetics did not provide any specific understanding of the RoF; it might be more insightful to compare the energies of chemically similar structures. To determine whether the \emph{magic} structures exhibit lower energy than structurally-similar \emph{non-magic} ones, we use Principal Covariates Regression (PCovR) \cite{pcovr, helfrecht_structure-property_2020},  a ML method which constructs a latent space projection to explore the correlation between stability and local symmetries within the dataset by expanding regression models to incorporate information on the structure of the input data, as implemented in the scikit-matter library\cite{skmatter, skmatter-ore}. 
In this mixing model, the projection is weighted towards the property of interest using a mixing parameter (of which a more extensive explanation is given in Section III of the SI), and, where the input linearly correlates with the target property, the resulting embedding will reflect this property along the first component, with subsequent components representing orthogonal dimensions in structure space.
In our case the PCovR is always trained on the species-tagged SOAP vectors and their formation energies.
We plot the first two principal covariates in Figure \ref{fig:pcovr_E}.
The first principal covariate is strongly correlated to the energetic descriptor, as can be seen in Figure \ref{fig:pcovr_E}, where in the lower plots we have colored each point in the projection by their \emph{magic} classification (left) and formation energy (right).
However, the second covariate (and all significant subsequent covariates, see Section IV of the SI) fail to separate the datasets into two distinct populations corresponding to this phenomenom.
This implies that for structurally similar compounds, there is no significant difference in energy between \emph{magic} and \emph{non-magic} samples.
We also see little difference in the spread of \emph{magic} versus \emph{non-magic} structures in the latent space, as shown by the kernel density probability map in the upper panel of Figure~\ref{fig:pcovr_E}.
Further principal covariates for the same PCovR representation are plotted in Section IV of the SI, as well as other relevant energetic descriptors (the energy above the convex hull energy, i.e., the envelope connecting the lowest energy compounds in the chemical space, and the band gap energy), in order to show how these targets yield similar results.  
Thus the RoF is neither correlated with the energetics, nor are \emph{magic} lower in formation energy when compared to chemically similar \emph{non-magic} ones.

\begin{figure}
    \centering
    \includegraphics[width=\columnwidth]{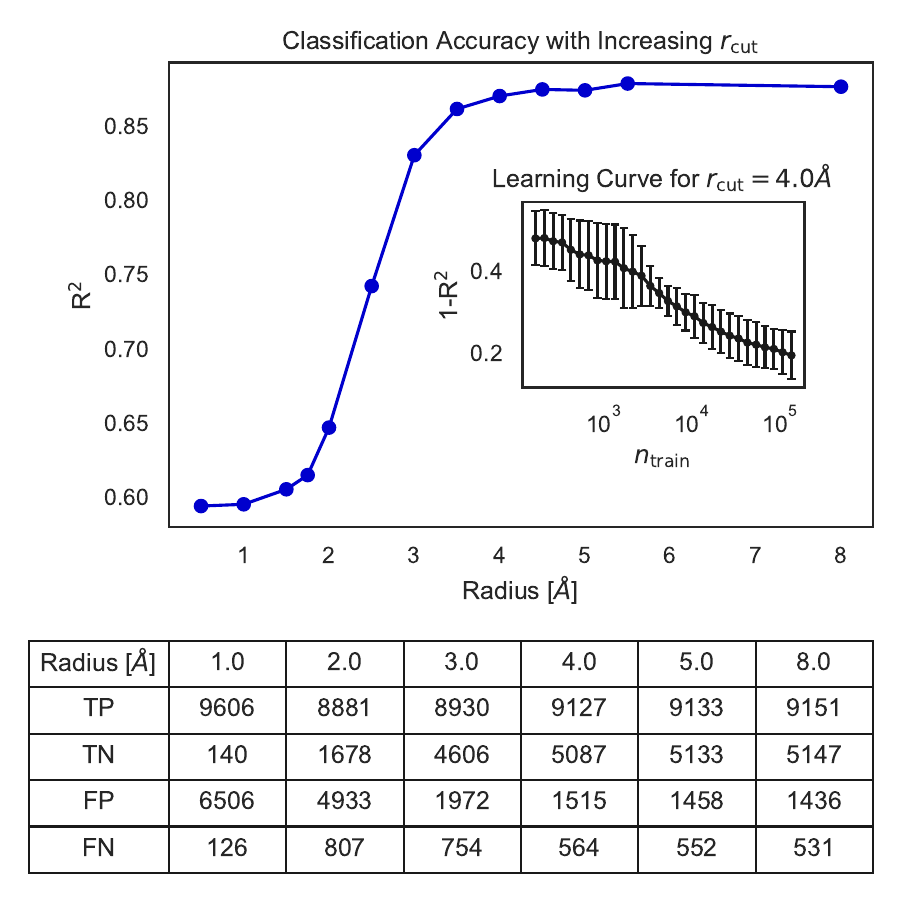}
    \caption{Random forest classification on local symmetries. Here we use the species-invariant 3-body SOAP vectors to build a random forest ensemble classifier.
    Accuracy saturates at approximately 4.0\AA, with little additional gain at larger cutoff radii.
    Below the figure we show the table of true positive (TP), true negative (TN), false positive (FP), and false negative (FN) results, showing that the classifier is unable to differentiate \emph{magic} and \emph{non-magic} structures at lower cutoff radii, leading to a high false positive (FP) rate.
    Inset in the upper figure is a learning curve for a cutoff radius of 4.0\AA, which shows a positive learning rate, albeit no saturation, an indication that secondary effects beyond the local environments play a role (or, more unlikely, that the dataset is not sufficiently large).
    }
    \label{fig:random-forest}
\end{figure}

The linear correlation between the average local symmetries and the RoF is not particularly strong (a logistic regression on the SOAP vectors results in an $R^2$ on the order of 0.6, as listed in Table 1 of the SI); thus, we turn to non-linear classifications to understand if the RoF is potentially correlated with these local neighborhoods.
We ignore the species information to focus solely on the average local symmetries.
We build a Random Forest (RF) classification\cite{breiman2001random} on both datasets, first varying the interaction cutoff that defines the local environment (see Fig.~\ref{fig:random-forest}).
We see a plateau in accuracy at 87\% as we consider local environments of 4.0\AA, suggesting that differentiating local symmetries occur within the first two neighbor shells, also supported by the high false positive (FP) rate at small cutoff radii.
From the learning curve on the 4.0\AA\  descriptors (inset), we see that the classification has a positive learning rate, although shows little saturation despite the large training set.
This result implies that local features are sufficient for the ML model to pick up the complexity of the datasets and to predict with good probability the correct classification.
We report the accuracy achieved by other classification algorithms in Section V of the SI.

\section{Conclusions}

Through an extensive investigation, in this work we highlight and analyze for the first time the anomalous abundance of inorganic compounds whose primitive unit cell contains a number of atoms that is a multiple of four -- a property that we name \textit{rule of four} (ROF) -- observed in both experimental and DFT-generated structure databases.
Here, we:
\begin{itemize}
    \item highlight the rule's existence, especially notable when restricting oneself to mostly experimentally known compounds; 
    \item explore its possible relationship with established energetic descriptors, namely formation energies, and utilise hybrid ML methods combining regression and principal component analysis to surprisingly rule out the possibility that the relative abundance has the (expected) effect of stabilising compounds, bringing them to a lower energy state; 
    \item conclude, through a global structural composition analysis of point groups and packing fractions, that the overabundance does not either correlate with high-symmetry structures, but rather to low symmetries and loosely packed arrangements maximising the free volume;
    \item predict, with an accuracy of 87\% the association to the rule of four of a compound by providing a random forest classification algorithm with local structural descriptors (the smooth overlap of atomic positions) only, eventually highlighting the importance of local symmetry rather than global one for the emergence of the \textit{rule of four}. 
\end{itemize}

This analysis constitutes a valuable reference for further systematic studies targeting the classification of materials’ features with novel ML approaches in order to screen for optimal experimental candidates.

% First, the study proves the existence of the anomaly by excluding a misclassification of \emph{magic} structures due to the unit cell primitivization procedure. 
% Second, an energetic analysis rules out that the relative abundance of \emph{magic} structures is driven by formation energies; hybrid ML methods combining regression and principal component analysis also allow to prove that the RoF does not have the expected effect of stabilizing compounds, bringing them to a lower energy state. A  point group and packing analysis concludes that the abundance of \emph{magic} structures within the datasets is rather correlated with low-symmetry configurations and loose packing effects, characterised by a relative low abundance of smallest to largest radii of elements constituting the compounds. 
% The role of local symmetry is offered by the RF classification, which is able to discern the phenomenon with an accuracy of 87\% using SOAP vectors containing the first-neighbor shell information; future studies studying the RoF should therefore focus on discerning which first-neighbor interactions are the most abundant ones and take into consideration the chemical local composition of each compound.

\section{Methods}

\subsection{Reduction to the primitive cell}

All the structures in both databases are reduced to the primitive cell using the \texttt{find\_primitive} function of the spglib \cite{spglib} package, varying the \texttt{symprec} value in the range of $1E-8$ to 1\AA.

\subsection{Scalar global descriptors}

% The structural geometry of compounds is mainly investigated through means of inherited symmetries, average coordination numbers and packing fractions' formalism.
% Due to the extensive nature of our datasets, we only briefly focus on chemical properties and relative elemental abundance, in favour of an in-depth geometrical approach, which is an easier framework to work on, given the structures' chemical diversity within the datasets.
% We are however aware of how structure and stability of inorganic solids are determined in general by short-range covalent and by long-range electrostatic forces \cite{kramer_relation_1991}, and that the symmetry of a structure depends sensitively on the balance between ionic and covalent forces \cite{borisov_geometrical_2000}.

The symmetry of compounds is investigated by looking at space groups and point groups.
% Of the total number of 230 space groups, groups 221--230 characterise structures with FCC symmetry, while groups 191--194 characterise structures with HCP symmetry \cite{dorst_point_2002}, which are the most closely packed geometries.
The point group of a given space group is the subgroup of symmetry operations over which the space group is invariant. With a total number of 32 point groups, it is easier to convey the symmetric properties of the vast variety of compounds via their point group rather than their space groups; while space groups uniquely identify geometric properties, point groups identify symmetry classes and reduce the parameter space to a lower degree when investigating the symmetries of all compounds.
The point groups are calculated through the \texttt{spglib}~\cite{spglib} and \texttt{seekpath}~\cite{seekpath} packages for the MC3D-source database, while we used the \texttt{SymmetryAnalyzer pymatgen} module -- which also relies on the \texttt{spglib} package developed by Togo and Tanaka~\cite{spglib} -- to find the symmetry operators and point groups for the MP dataset.
As concerns packing mechanisms, we extend the conventions employed by Hopkins\cite{Hopkins_2011} to $n$-elements packing and employ the $\alpha$, $PF$ and $x$ parameters.
In structures with FCC and HCP symmetry, the maximum packing fraction is 0.74. 
$\alpha$=1 denotes unary compound.
Conversely, when $\alpha$ $\sim 0$ the compounds contain elements whose radii distribution presents a wider spread.

% Local rearrangements within structures can also be analyzed by means of coordination numbers.
% The coordination number of an atom~$j$, $CN_{j}$, represents the number of atoms located within a sphere of radius $r_{cut}$ centred around $j$, which are at a distance $r_{ij}$ from $j$.
% Expressed analytically, 

% \begin{equation}
%     CN_j=\sum_{i\ne j}f(r_{ij}), \hspace{.5 cm} f(r_{ij}) = \begin{cases}
%       0, & \text{if}\ r_{ij} \leq r_{cut} \\
%       1, & \text{if}\ r_{ij}> r_{cut}
%     \end{cases} ~~~.
% \end{equation}

% The cut-off distance $r_{cut}$ is the minimum distance between the atoms in the cluster.

% The pymatgen \cite{pymatgen} class CrystalNN, based on a Voronoi \cite{urusov_terms_2014, Damasceno453, waroquiers_statistical_2017} algorithm which uses the solid angle weights to determine the  most probable coordination environment for each atom, is used to calculate the average coordination number.
% Neighbor distances greater than the covalent radii sums are enforced to have zero weight and the $r_{cut}$ is adjusted for each situation with the Voronoi algorithm.

\subsection{Local symmetry descriptors and ML pipeline}

We adopt the following ML pipeline to study local symmetries and energetic effects.
First, the atomic \emph{representation} of each compound is obtained with SOAP vectors (see section II of the SI), computed with the \texttt{librascal} library \cite{musil_efficient_2021}.
The SOAP features are then averaged within each compound, and the representations from the two datasets are normalised simultaneously. 
We then select a diverse subset of 2\,000 features through Furthest Point Sampling (FPS) algorithm \cite{eldar_farthest_1997, cersonsky_improving_2020, skmatter, skmatter-ore} (see Section II of the SI), efficiently reducing the dataset size without losing important information. 
For Sec.\ref{subsec:soap}, we perform a linear ridge regression with 4-fold cross-validation -- which optimises the regularisation parameter to prevent overfitting -- on the formation energies data retrieved from the MP database to ascertain the accuracy of the model.
Table \ref{table:pcovr_mp_tab} illustrates the RMSE and the accuracy in units of eV of the predicted energetic quantities.

\begin{table}[htbp!] 
\centering
\begin{tabular}{|p{4.6cm}|p{1.3cm}|p{1.9cm}|}
\hline
Predicted quantity & RMSE & Uncertainty \\
\hline
\hline
Formation Energy per atom (eV) &  0.0530 & 0.4002 eV  \\
\hline
Energy above Convex Hull per atom (eV) & 0.2938 & 4.0006 eV  \\
\hline
Band Gap Energy with PBE-DFT functional (eV) & 0.3097 & 3.6560 eV \\
\hline
\end{tabular}
\vspace{0.2 cm}
\caption{
    RMSE and uncertainty in units on the predicted energetic quantities for the MP database.
    The ML algorithm is a LRR with a 4-fold cross-validation.
    We report the formation energy per atom, the energy above the convex hull and the band gap energy.
}
\label{table:pcovr_mp_tab}
\end{table}

Compared to results in the literature, which achieve an accuracy in formation energy prediction of 0.173 eV (Automatminer \cite{dunn_benchmarking_2020}) and 0.0332 eV (Crystal Graph Convolutional Neural Networks\cite{ziletti_insightful_2018}), the accuracy of  0.4002 eV is sufficient for this study, since the aim of our study is not to find the most efficient way to predict energies, but rather to provide a sufficient regression prediction to employ in PCovR analysis.
We use the species-invariant SOAP vectors to \emph{classify} the RoF phenomenon using \texttt{scikit-learn}'s \cite{scikit-learn} \texttt{RandomForestClassifier} algorithm \cite{random_forest}, which accepts binary labels as target properties (\emph{magic} or \emph{non-magic}) and outputs a probability between 0 and 1 for each compound to fall into the \emph{magic} subset.
Training and testing set constitute respectively 90 and 10\% of the whole dataset.
Our random forest classification comprises 100 random decision trees. 
This classifier performs better in our case compared to Support Vector Machine (SVM) and Logistic Regression (LR) classifiers, signifying a need for a stochastic model.

% When the stability of compounds is analysed in terms of their energetic descriptors, the initial hypothesis correlating stability and \emph{magic} abundance is again disproved, as can be seen from the results in Figure~\ref{fig:pcovr_E}.
% The insertion of energetic descriptors into the PCovR model does not classify the problem.
% However, it is interesting to notice how PCovR plots of the principal covariates within the MP database offer some qualitative insights supporting the histograms in Figure~\ref{fig:packing}.
% When $\beta$ is lowered to 1, only maintaining the SOAP vectors PCA representation, some agglomerates of O-Mg compounds with right-angle bonds and low formation energy emerge, albeit this group is only a small subset of the \emph{magic} structures.
% When, instead, the PCovR (with $\beta$=0.5) is tailored to contain SOAP vectors and geometric descriptors, all combined together in one n-dimensional array, some clustering emerges representations, as illustrated in Figure~\ref{fig:mp_geo}.
% It is interesting to notice that the Mg-O square bonds still appear, validating the SOAP representation's usefulness in separating the specific subgroup.
% The top panel on the right of the figure is in agreement with the consideration relating Mg-O square bonds with low formation energy, being them characterized by a high $PF$.

\section{Data Availability}
The full dataset employed for the analysis can be downloaded from the Materials Cloud Archive \cite{materialscloudarchive}, where the MC3D-source data is only provided in SOAP format as the experimental structures can not be released due to licensing constraints. Its DFT--relaxed counterpart is available at: \href{https://archive.materialscloud.org/record/2022.38}{https://archive.materialscloud.org/record/2022.38}. Instead, we provide the full list of structure IDs for each database, including the version of the database upon the time of extraction. 

\section{Code Availability}
The codes to reproduce the results and figures can be found at: \href{https://github.com/epfl-theos/r4-project}{https://github.com/epfl-theos/r4-project}. As the MC3D-source structure data cannot be made publicly available due to licensing contraints, the repository contains example data from a reduced random subset of the publicly available MP dataset in order to test run a preliminary analysis.  

\section{Acknowledgements}
This work was supported by a MARVEL INSPIRE Potentials Master's Fellowship
and the MARVEL National Centre of Competence in Research (NCCR) funded by the Swiss National Science Foundation (grant agreement ID 51NF40-182892).
RKC acknowledges funding from the European Research Council (ERC) under the European Union's Horizon 2020 research and innovation programme under grant agreement No 677013-HBMAP.
MB acknowledges funding by the European Union’s Horizon 2020 research and innovation program under grant agreement No. 824143 (European MaX Centre of Excellence “Materials design at the Exascale”).
CSA acknowledges funding from the European Union's Horizon 2020 research and innovation programme under grant agreement No 760173.
The  authors  thank  the AiiDA team for giving access to the full MC3D-source dataset, and are grateful to the technical and theoretical insights offered by the members of MARVEL, THEOS, and COSMO.

\section{Author contributions}
E.G. conducted the analysis on the data and wrote the paper in collaboration with C.S.A and R.K.C.
R.K.C. instructed on how to use many of the employed computational tools, designed the computational strategy, and helped conduct the analysis and interpretation of the data.
M.B. provided technical support, especially on the MC3D-source dataset and helped with the analysis and interpretation of the data.
C.S.A. managed the project, wrote the paper, and helped with the analysis and interpretation of the data.
N.M. supervised the project, helped with the interpretation of the results, and provided the initial suggestion for the investigation.
All authors edited and reviewed the paper.

\section{Competing Interests}
There are no competing interests to declare.

\section{Additional Information}
Supplementary Information is available for this paper.

% \printbibliography
\bibliography{bib.bib}
\end{document}

\maketitle

\newpage

\section*{I. Materials Databases}

In this section we introduce the two datasets employed in the study and explain how the raw data is obtained.

\subsection*{The Materials Cloud 3-dimensional crystals Database - MC3D}

The MC3D~\cite{mc3d} is a database of structures optimized with the Quantum ESPRESSO code~\cite{Giannozzi2009, Giannozzi2017} using fully-automated workflows developed in AiiDA~\cite{aiida, huber_aiida_2020}.
The starting set of structures for the geometry optimization is obtained from the COD~\cite{COD-ref}, the ICSD~\cite{icsd_n} and the MPDS~\cite{Pauling} databases.
Each CIF file is parsed via an AiiDA workflow that removes unnecessary tags, performs minor corrections to the syntax, and parses the contents to extract the corresponding structure.
The parsed structures are subsequently normalized and primitivized using SeeK-path~\cite{seekpath}, and a uniqueness analysis is performed to remove duplicate structures.
Finally, hydrogen-containing structures from the COD are removed due to the prevalence of molecular crystals in this database, and any structure containing an actinide is also excluded from the database.
The resulting 79\,854 structures \textit{before geometry optimization} are labelled as MC3D-source and used for the analysis in this paper.
% Most (63\,093) of the structures in the MC3D-source come from the MPDS, 13\,798 were obtained from the ICSD and 2\,963 from the COD.
In this early version of MC3D-source, most (63 093) of the structures came from the MPDS, 13\,798 were obtained from the ICSD and 2\,963 from the COD.
Although the vast majority of the structures in the MC3D-source are experimental, some of the structures extracted from the ICSD and COD were found to be flagged as theoretical, i.e. hypothesized in a theoretical study instead of being observed experimentally.
Screening the metadata for these flags, we find 3\,071 theoretical structures, so approximately 3.85\% of the full structure set.

Due to licensing constraints, we are not allowed to publish the full MC3D-source structure set.
Instead, we provide a YAML file on the Materials Cloud archive~\cite{materialscloudarchive} called \texttt{MC3D\_ids.yaml} that contains the list of versions and IDs for each structure extracted from the three databases.

\subsection*{Materials Project - MP}

The Materials Project\,(MP) \cite{jain_commentary_2013} dataset used contains a total of 83\,989 bulk, crystalline, inorganic compounds that have been relaxed with first-principles calculations starting from experimental databases or from structure-prediction methods. It is retrieved through the Matminer ~\cite{matminer} Python library. The version of the database employed in the study dates back to 10/18/2018, corresponding to the \texttt{p\_all\_20181018} dataset retrieved with the \texttt{matminer.datasets} module~\cite{data_access}.

\newpage

\section*{II. SOAP and FPS} 
In this study, it is necessary to have an ML representation that is invariant to symmetry operations and changes smoothly with the Cartesian coordinates.
We choose Smooth Overlap of Atomic Positions (SOAP) vectors, a representation based on smoothed atomic densities: these are abstract feature vectors based on an expansion of atom-centered Gaussians in radial basis functions and spherical harmonics.
This representation discretizes a three-body correlation function including information on each atom, its relationships with neighbouring atoms, and the relationships between sets of neighbours, quantifying similarities between atomic neighborhoods. 

% Using a kernel, the originally linear operations of PCA are performed in a reproducing kernel Hilbert space (RKHS). 
The 3-body SOAP vector is built as 
\begin{equation}
\braket{\alpha n \alpha^\prime n^\prime l |\mathcal{X}} \propto \frac{1}{\sqrt{2l+1}}\sum_m \braket{\alpha n lm |\mathcal{X}}^*\braket{\alpha^\prime n^\prime lm |\mathcal{X}}
    % \rho_{i}^{\alpha}(r)=\sum_{j\in \alpha}N_{\sigma}(r-r_{ij})=\sum_{nlm}\langle \alpha n l m | X_{i} \rangle B_{nlm}(r) ~~~, 
\end{equation}
where $\alpha$ refers to the species of the considered atoms, and $\braket{\alpha n lm |\mathcal{X}}$ is the expansion of a density field over spherical harmonics and radial bases with $n$ radial bases and $l$ angular channels
\begin{equation}
    \braket{\alpha n lm |\mathcal{X}} = \int d\mathbf{r} R_n(r) Y_m^l(\hat{\mathbf{r}})\braket{\alpha r |\mathcal{X}}
\end{equation}
Here $\braket{\alpha r |\mathcal{X}}$ encodes the species-tagged density field as a function of $r$ with radius $\sigma$ and accumulated until an interaction cutoff of $r_{cut}$.
For the examples contained in the text, we have used $n_{max} = 4$, $l_{max} = 4$, $\sigma = 0.5$, and $r_{cut} = 3.5$. When using species-invariant SOAP vectors, we use the same hyperparameters but combine all species channels together.

When working with species-tagged SOAP vectors, in order to increase computational efficiency we select the most diverse features using a Furthest Point Sampling (FPS) algorithm, an unsupervised selection method which maximizes diversity (variance) of the selected vectors as measured by the mutual Euclidean distance.

\newpage

\section*{III. PCovR: tuning the mixing parameter} 

The Principate Covariates Regression (PCovR) \cite{helfrecht_structure-property_2020} combines the losses of Linear Ridge Regression (LRR) and Principal Component Analysis (PCA) through the mixing parameter $\beta$. 
The feature matrix which embeds the reduced SOAP representation is projected into latent space with an orthogonal projection. 
Finding the optimal projection to the latent space amounts to minimizing the loss, which happens when the projection is built out of the principal eigenvectors of the covariance matrix of the initial feature matrix.

\begin{figure*}[htbp!]  
    \centering
    \includegraphics[width=1\columnwidth]{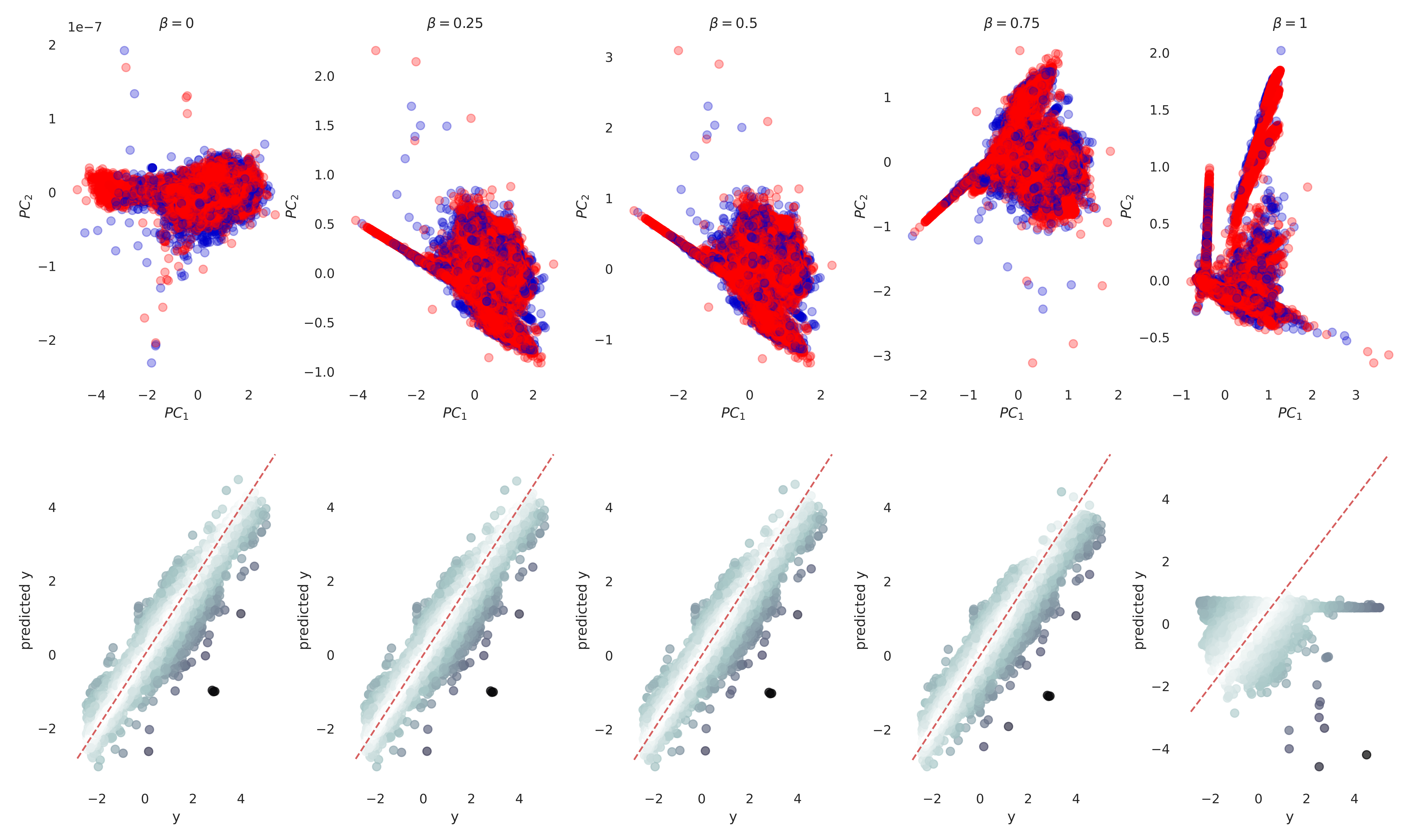}
  \caption{The combination of LR (far left) and PCA (far right) in the PCovR analysis on the MP database. The resulting projections and regressions are shown at the indicated $\beta$ values. \textit{Magic} compounds are coloured in red, while \textit{non-magic} ones in blue.} \label{fig:pcovr_change}
\end{figure*}

\newpage

\section*{IV. Energetic analysis with PCovR} 

\begin{figure*}[htbp!] 
\centering
\includegraphics[width=.45\columnwidth]{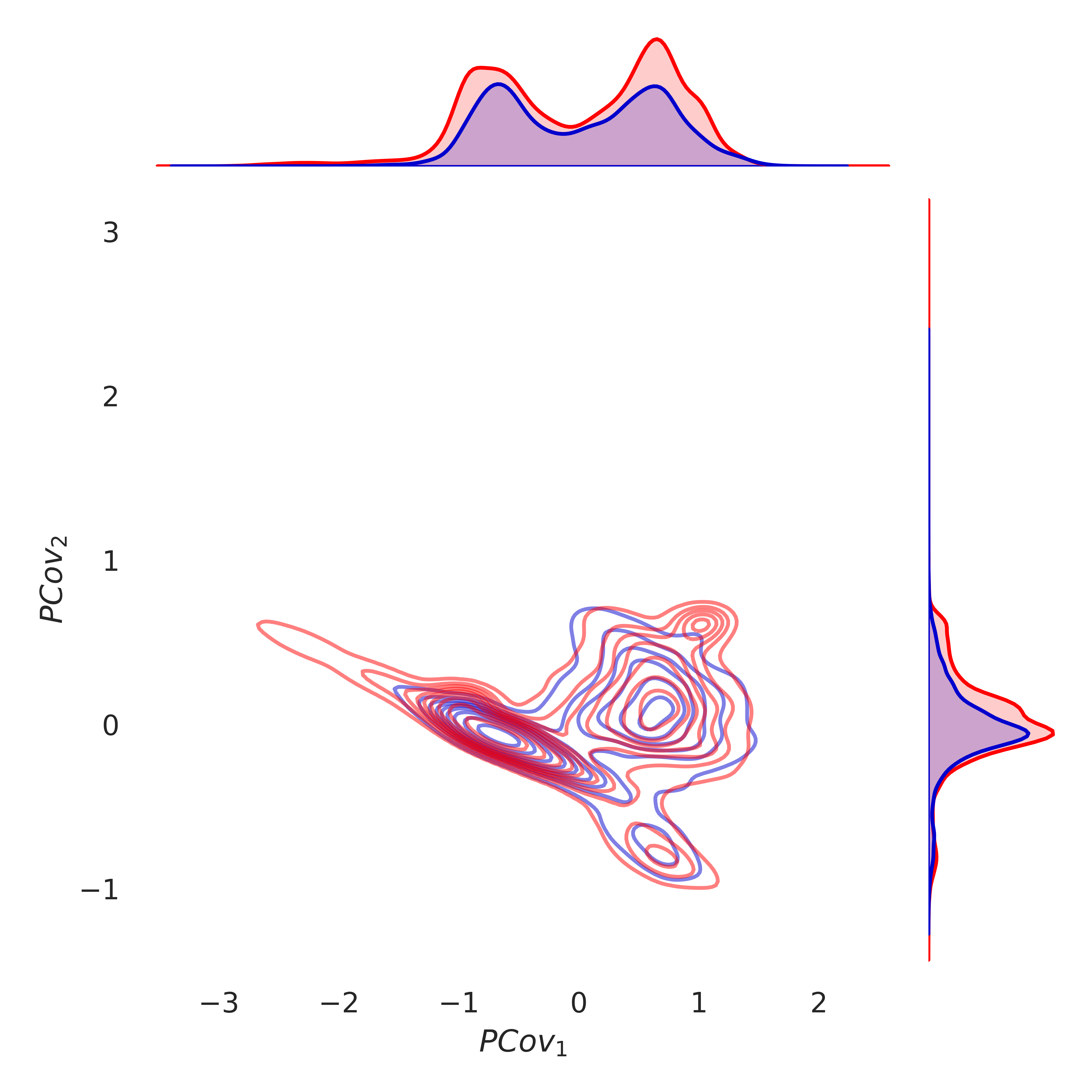}
\includegraphics[width=.45\columnwidth]{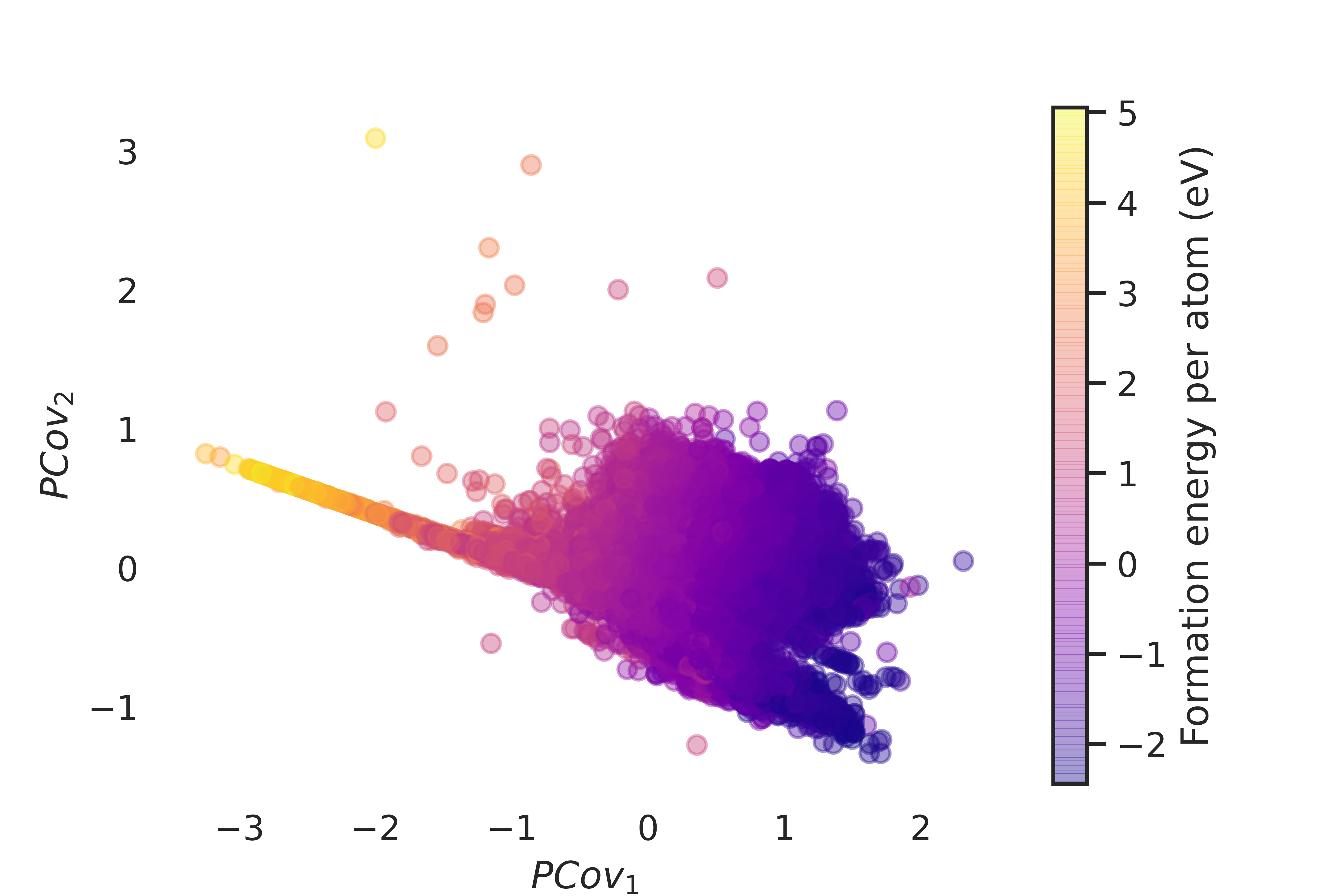}
\includegraphics[width=.45\columnwidth]{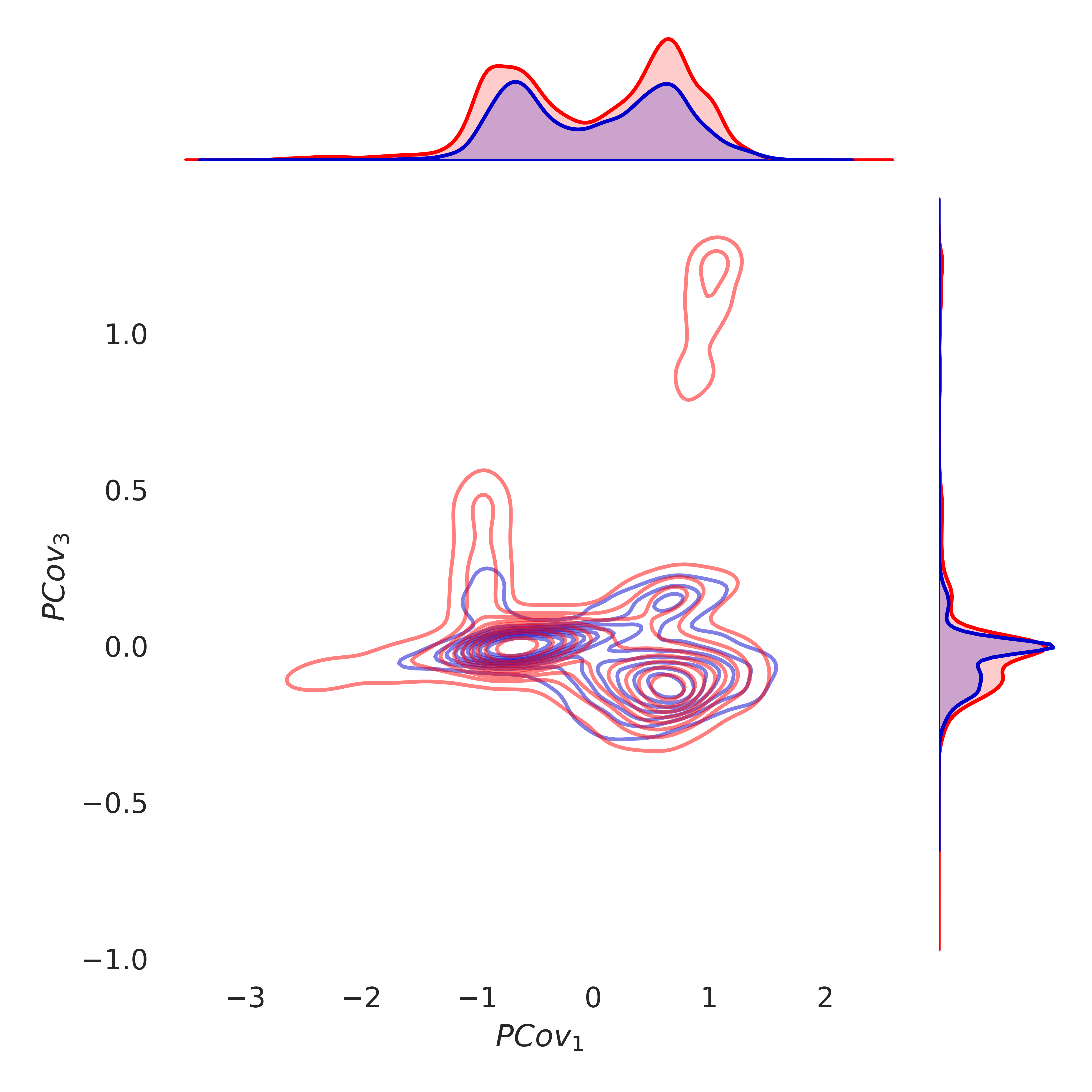}
\includegraphics[width=.45\columnwidth]{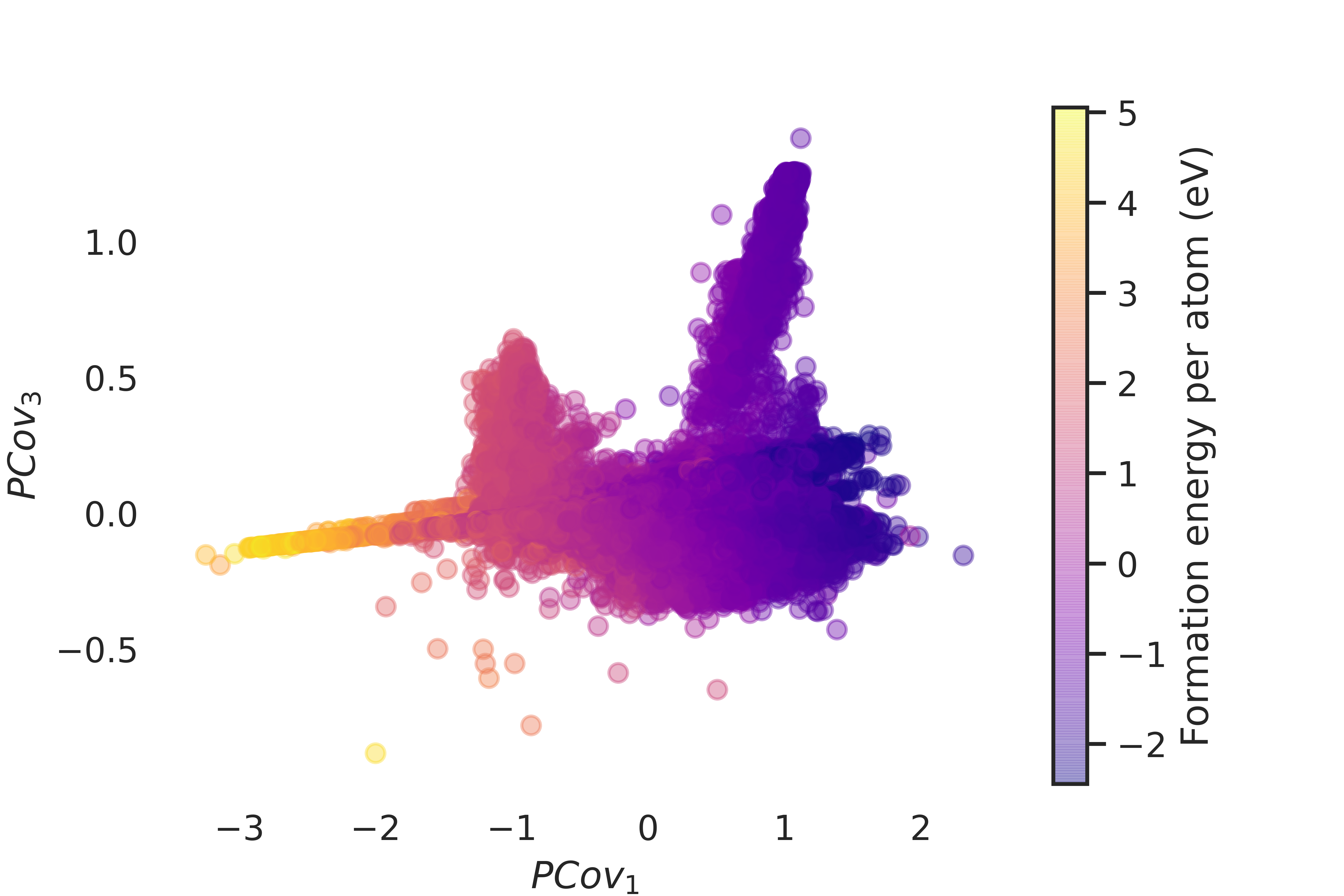}
\includegraphics[width=.45\columnwidth]{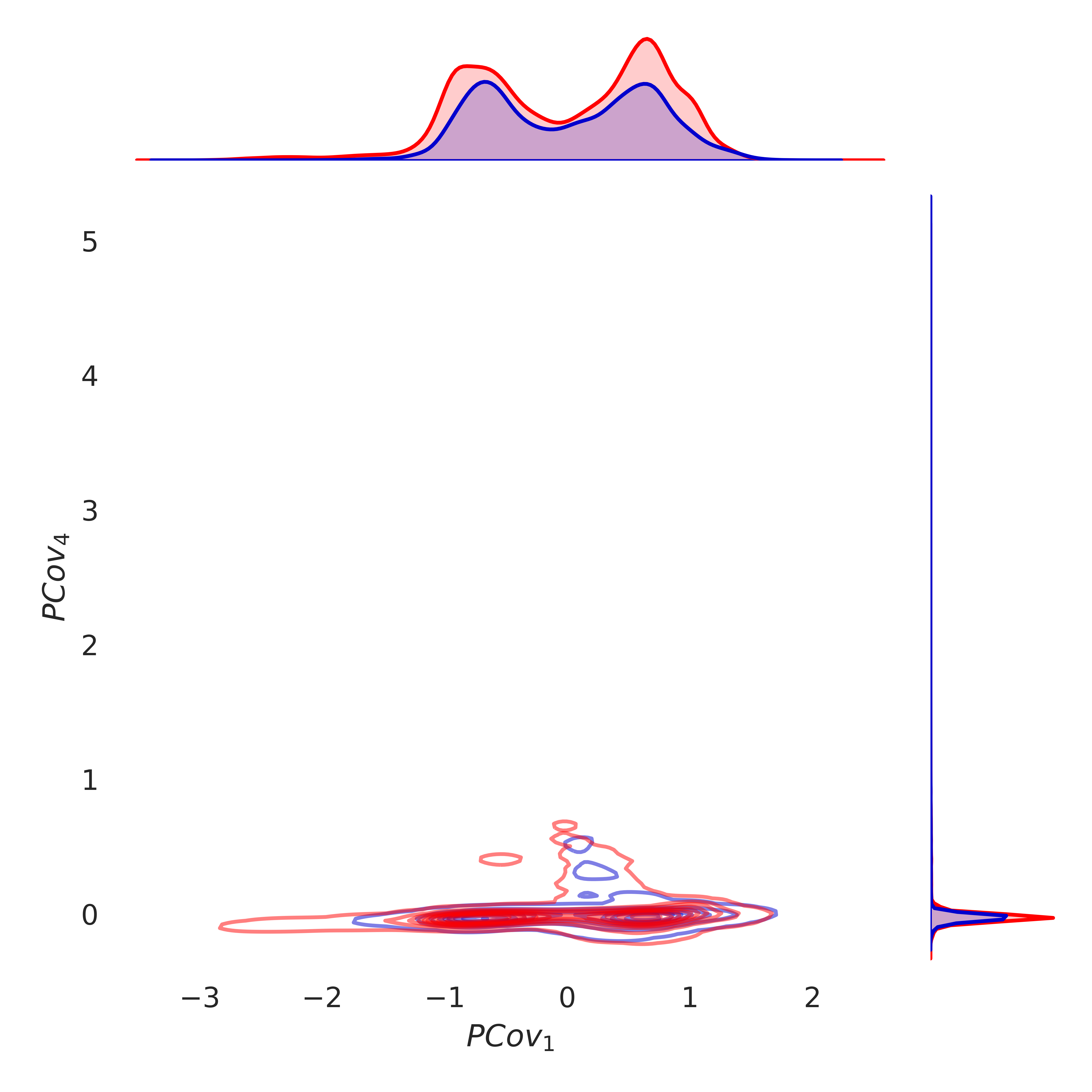}
\includegraphics[width=.45\columnwidth]{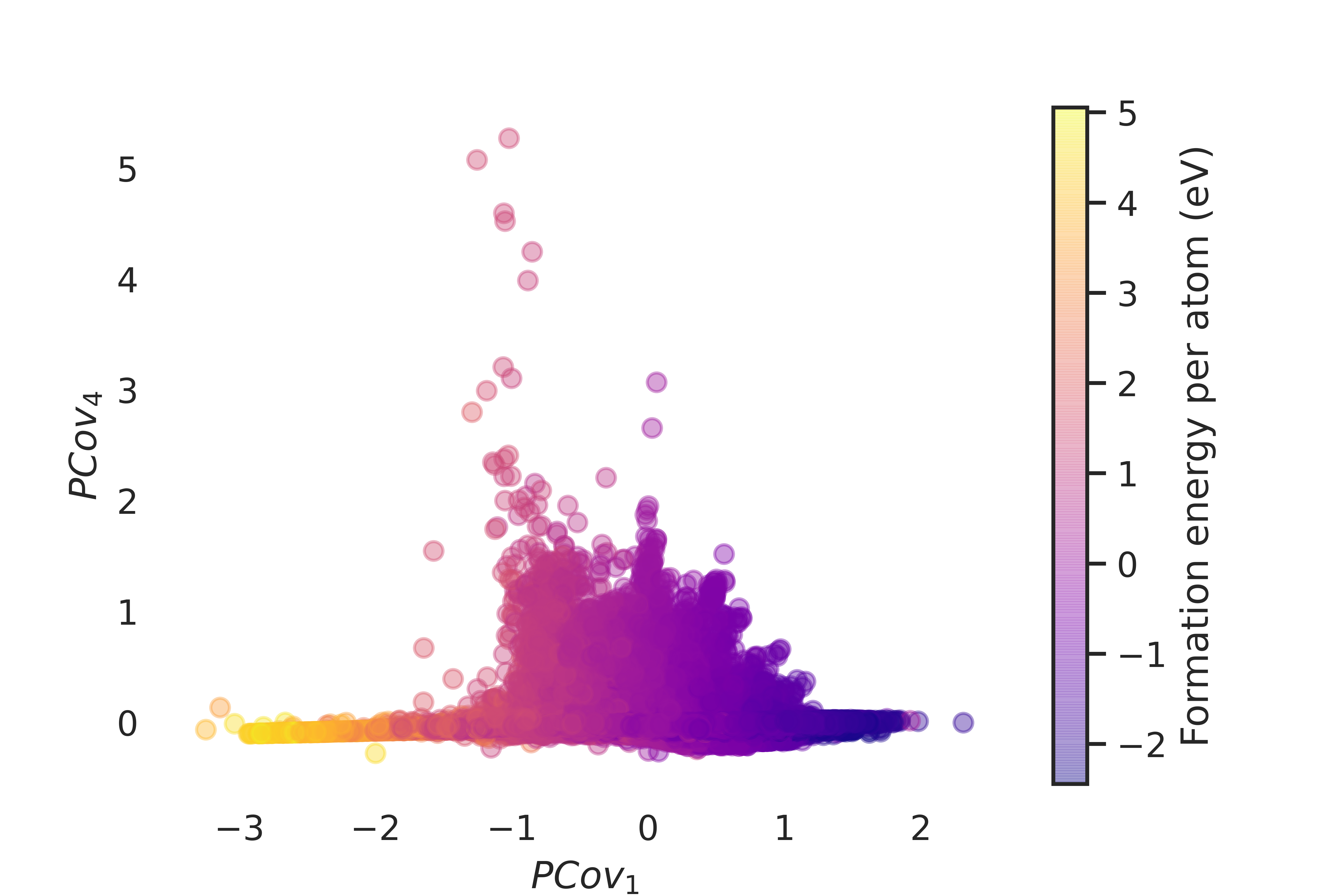}
\caption{PCovR representation with $\beta$ = 0.5 containing information on SOAP representation and regressed on the formation energy per atom. }
\label{fig:e_form}
\end{figure*}

\begin{figure*}[htbp!] 
\centering
\includegraphics[width=.45\columnwidth]{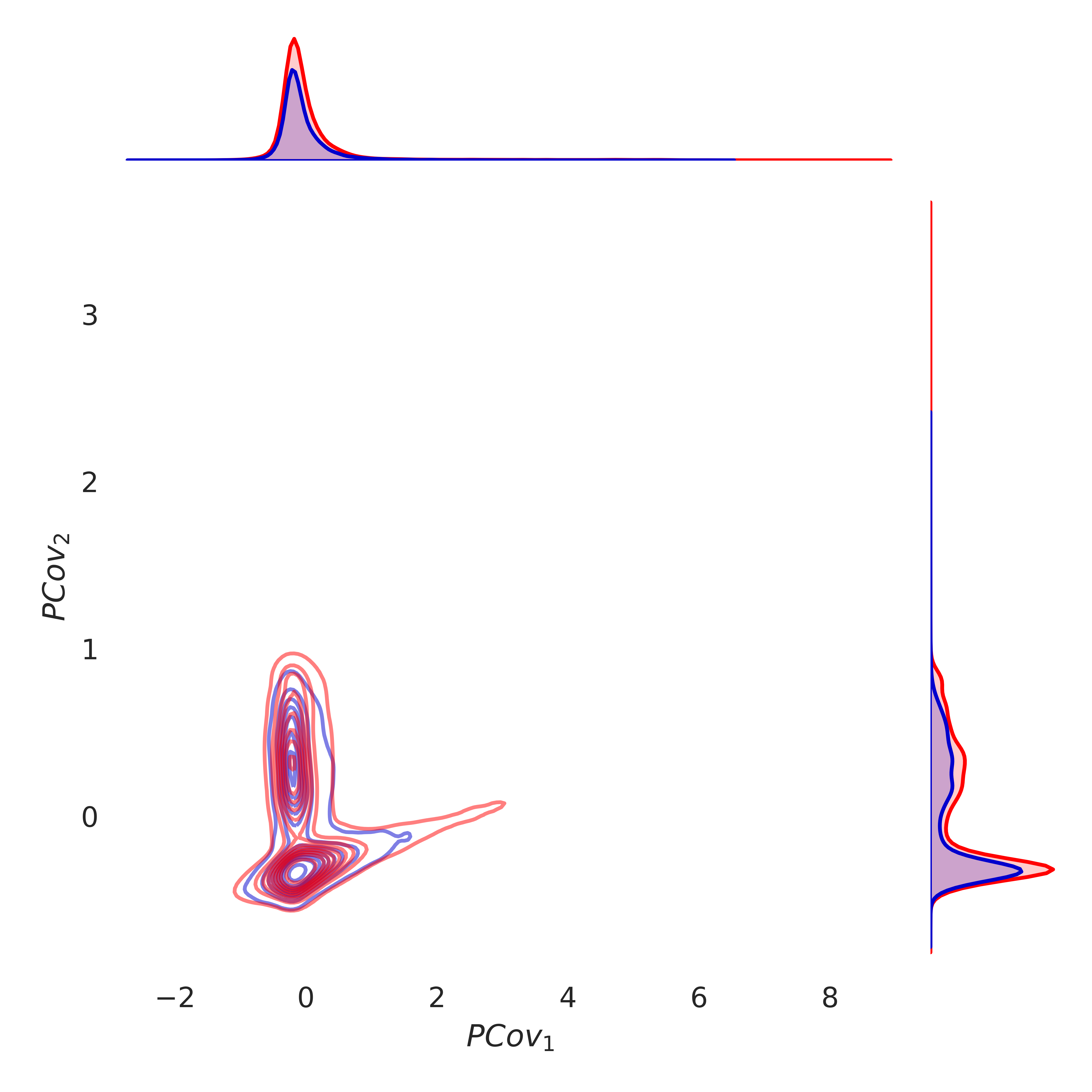}
\includegraphics[width=.45\columnwidth]{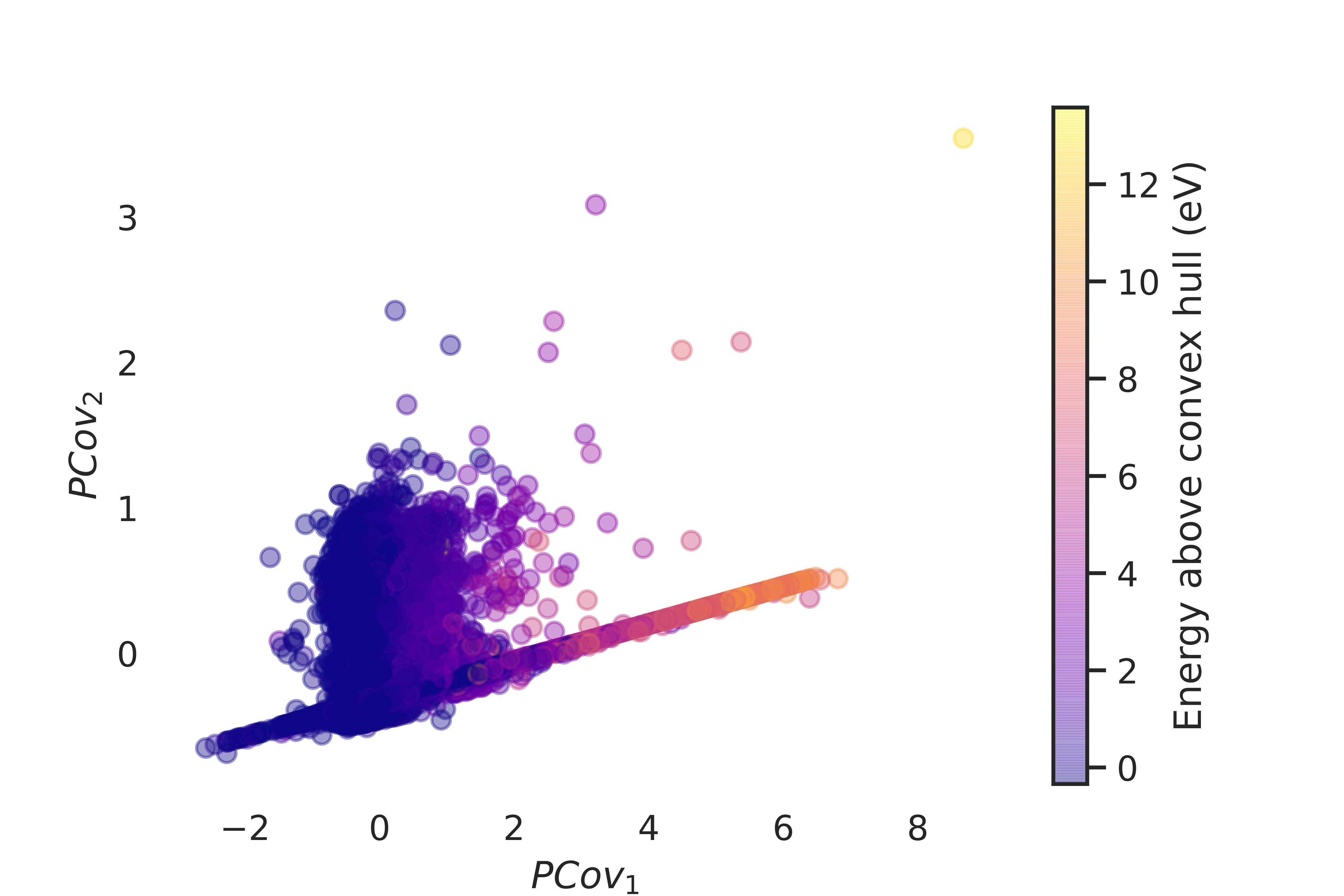}
\includegraphics[width=.45\columnwidth]{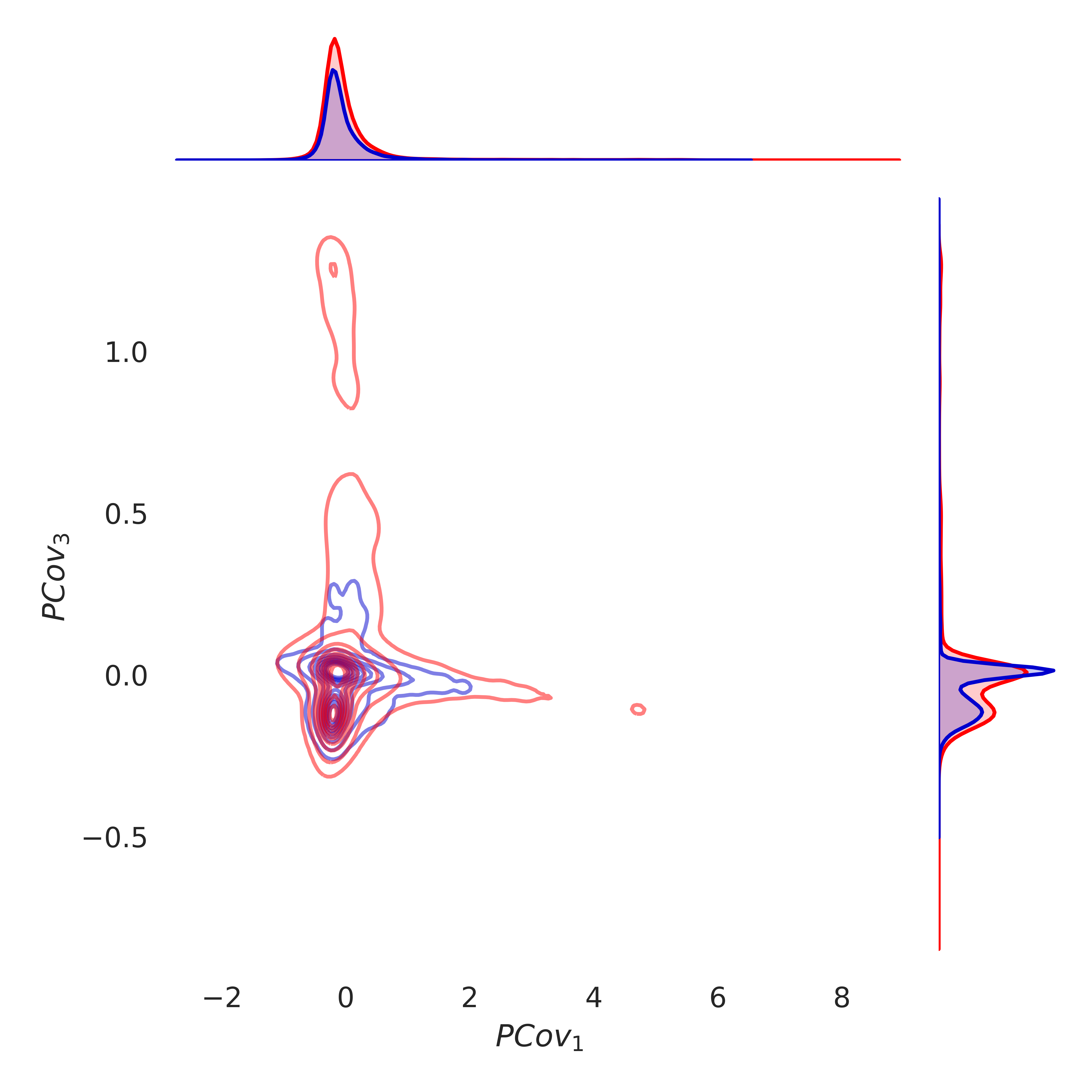}
\includegraphics[width=.45\columnwidth]{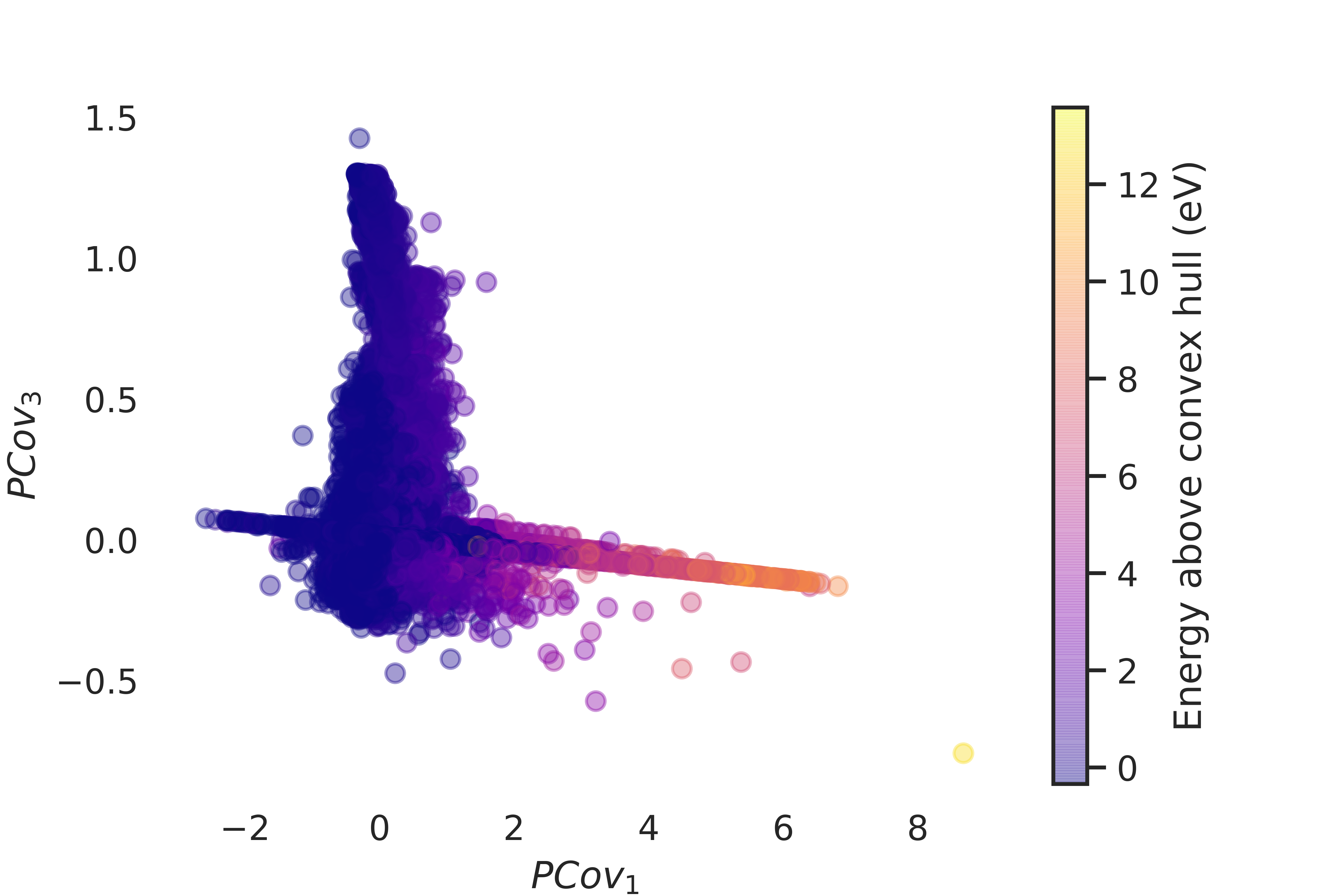}
\includegraphics[width=.45\columnwidth]{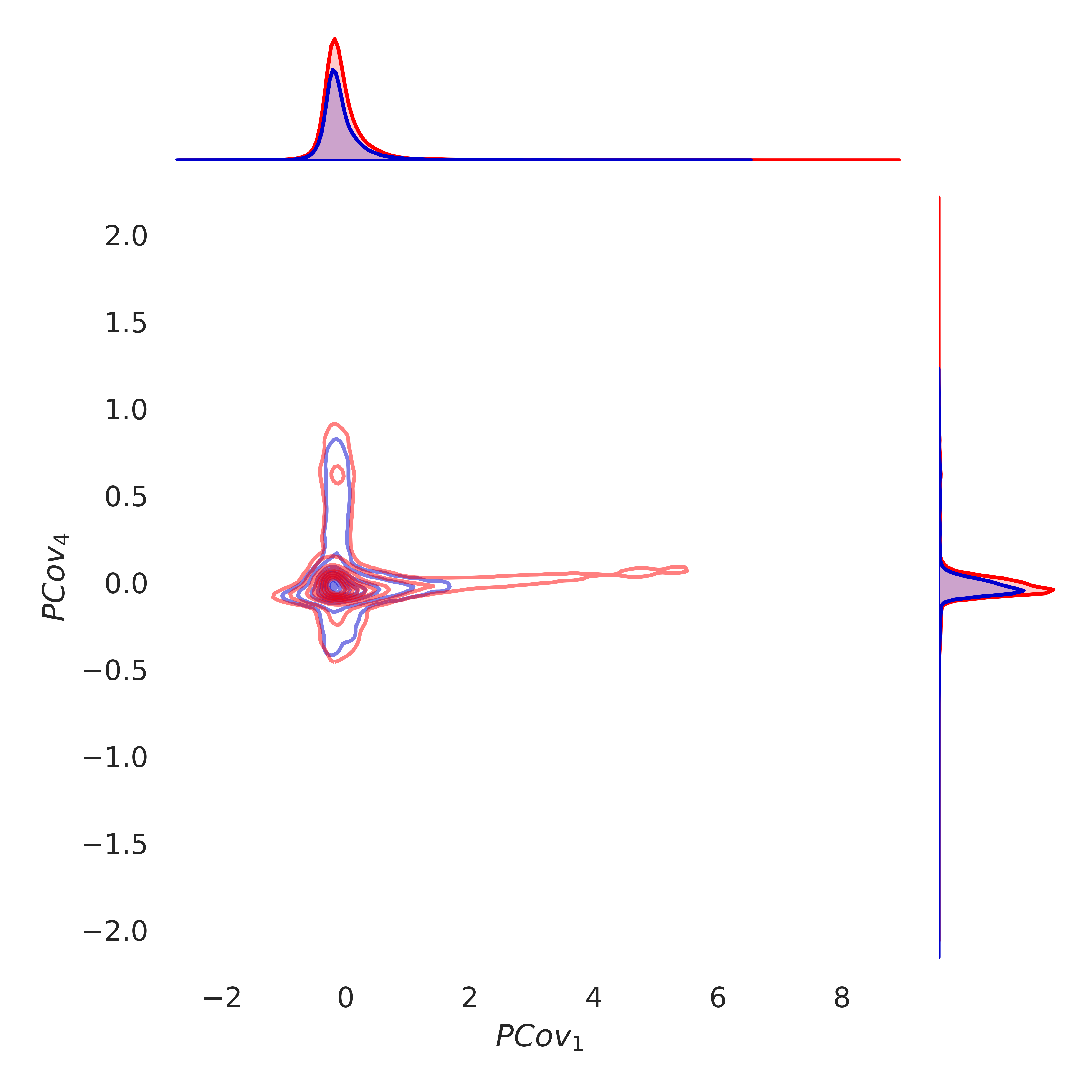}
\includegraphics[width=.45\columnwidth]{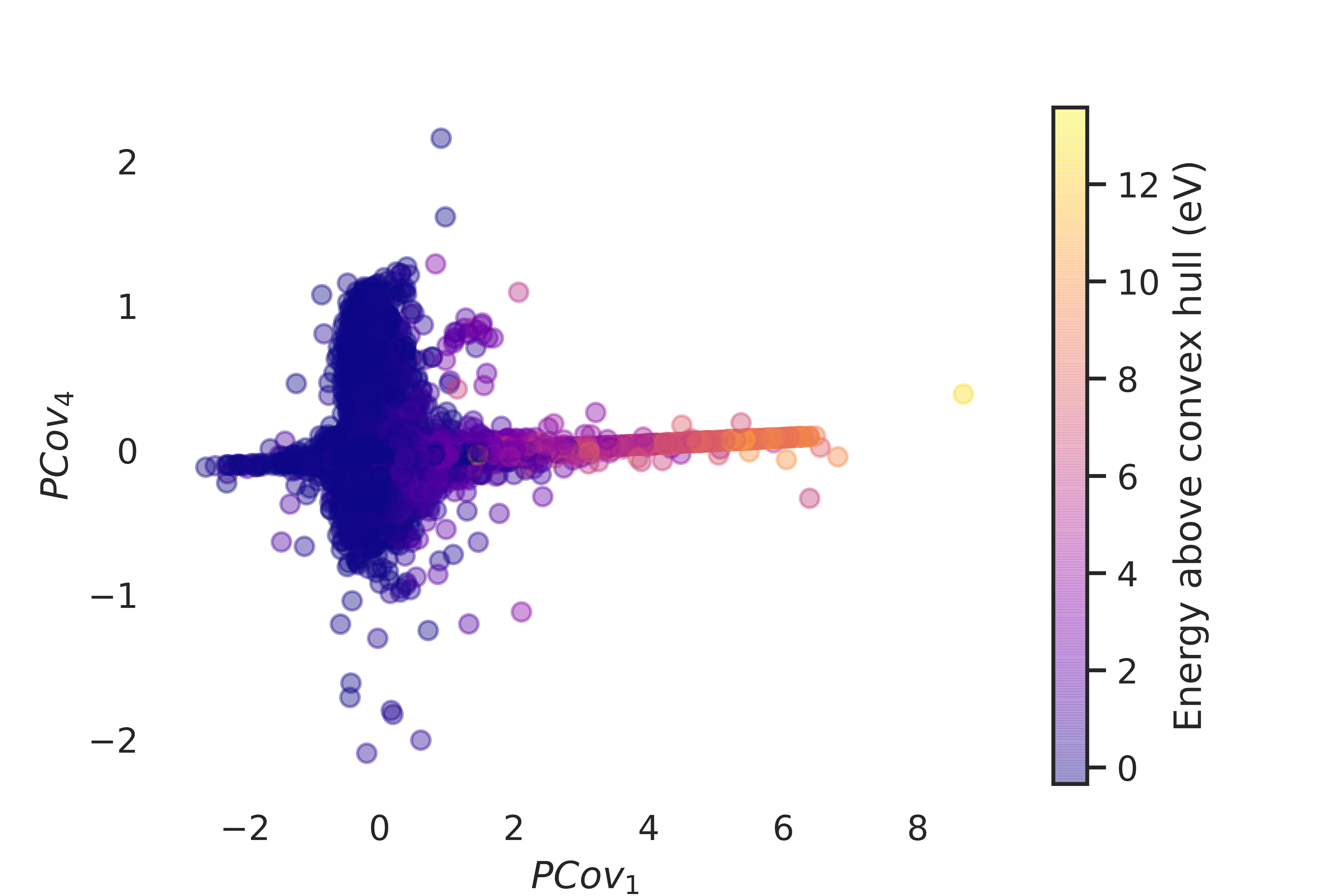}
\caption{
    PCovR representation with $\beta$ = 0.5 containing information on SOAP representation and regressed on the energy above the convex hull. }
\label{fig:convex_hull}
\end{figure*}

\begin{figure*}[htbp!] 
\centering
\includegraphics[width=.45\columnwidth]{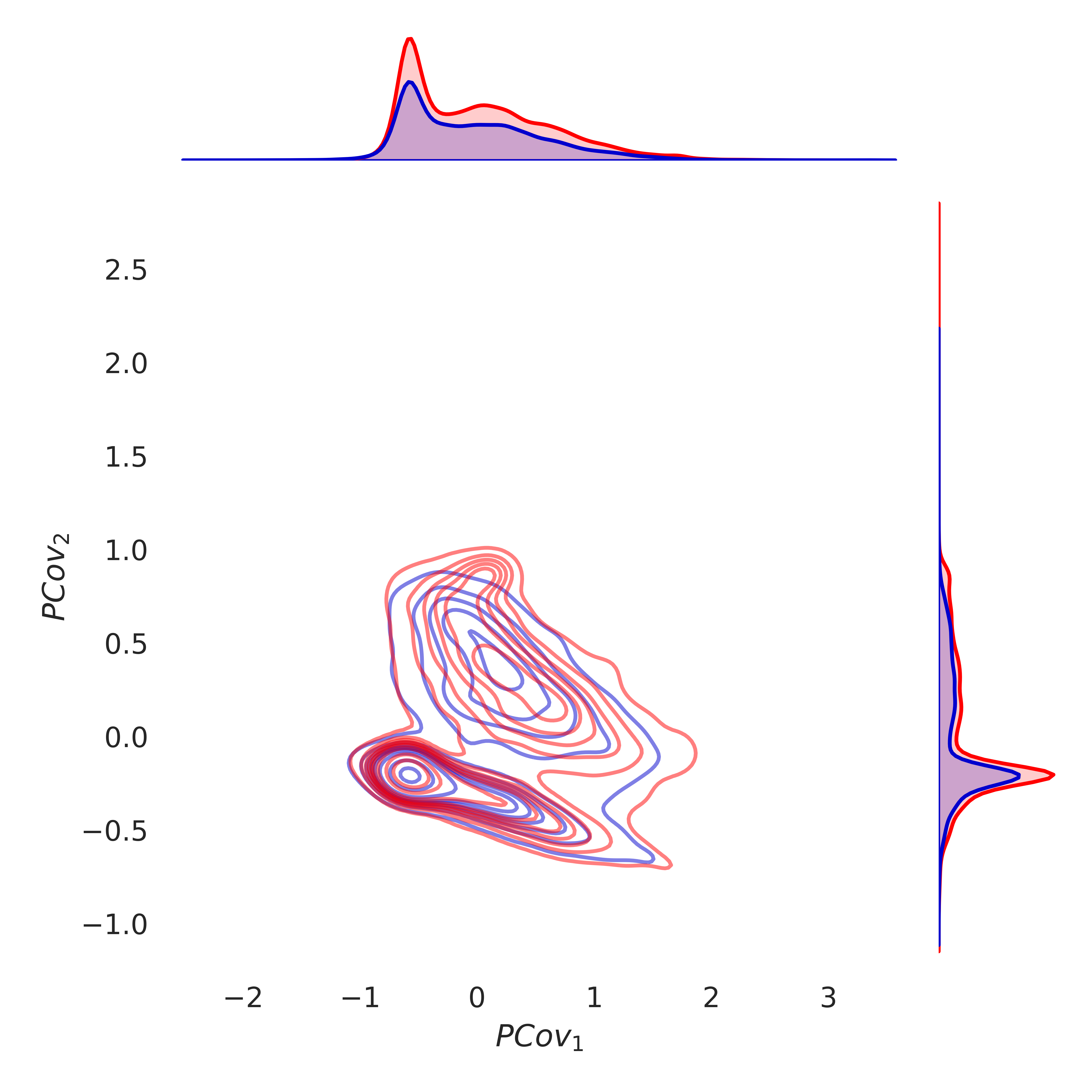}
\includegraphics[width=.45\columnwidth]{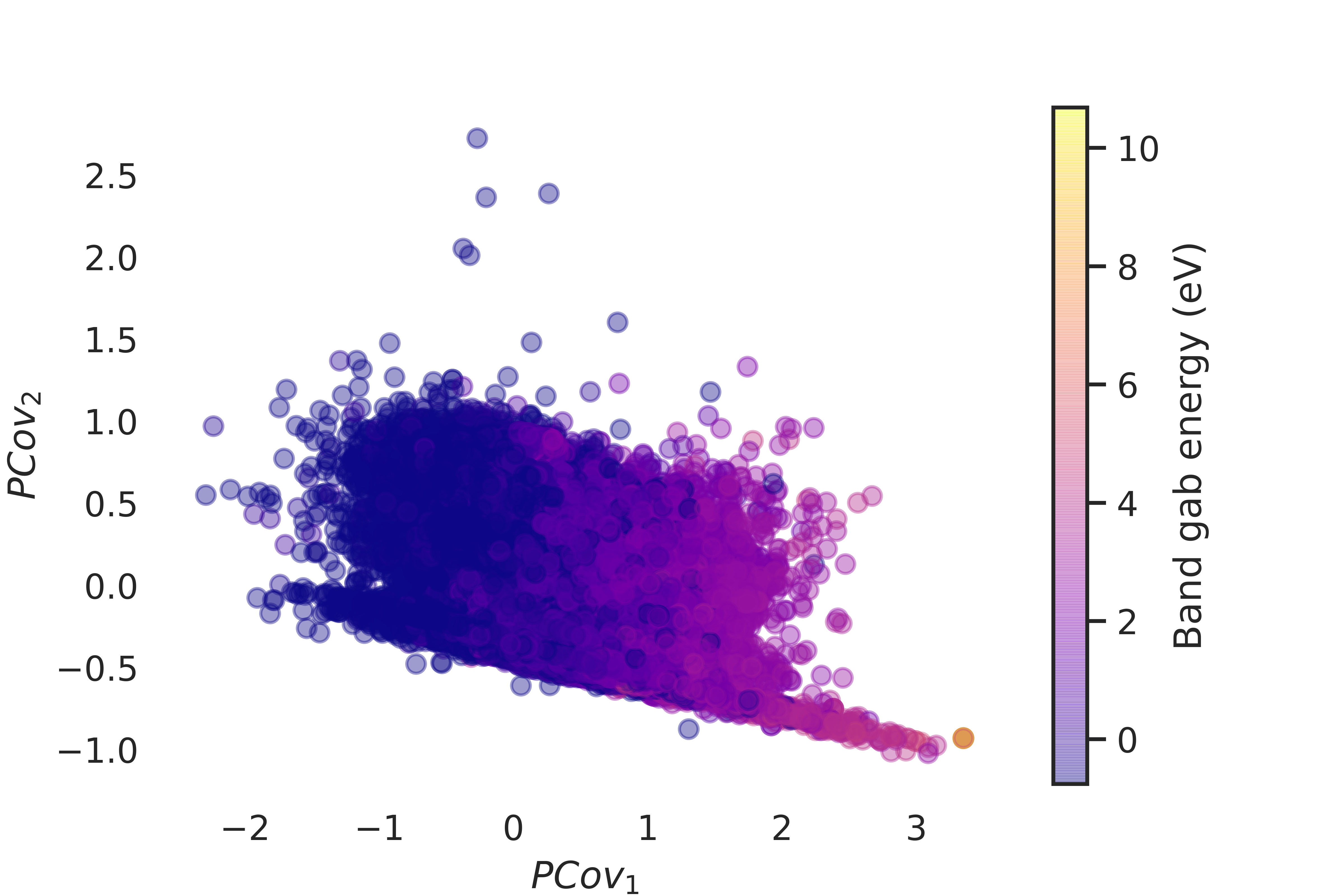}
\includegraphics[width=.45\columnwidth]{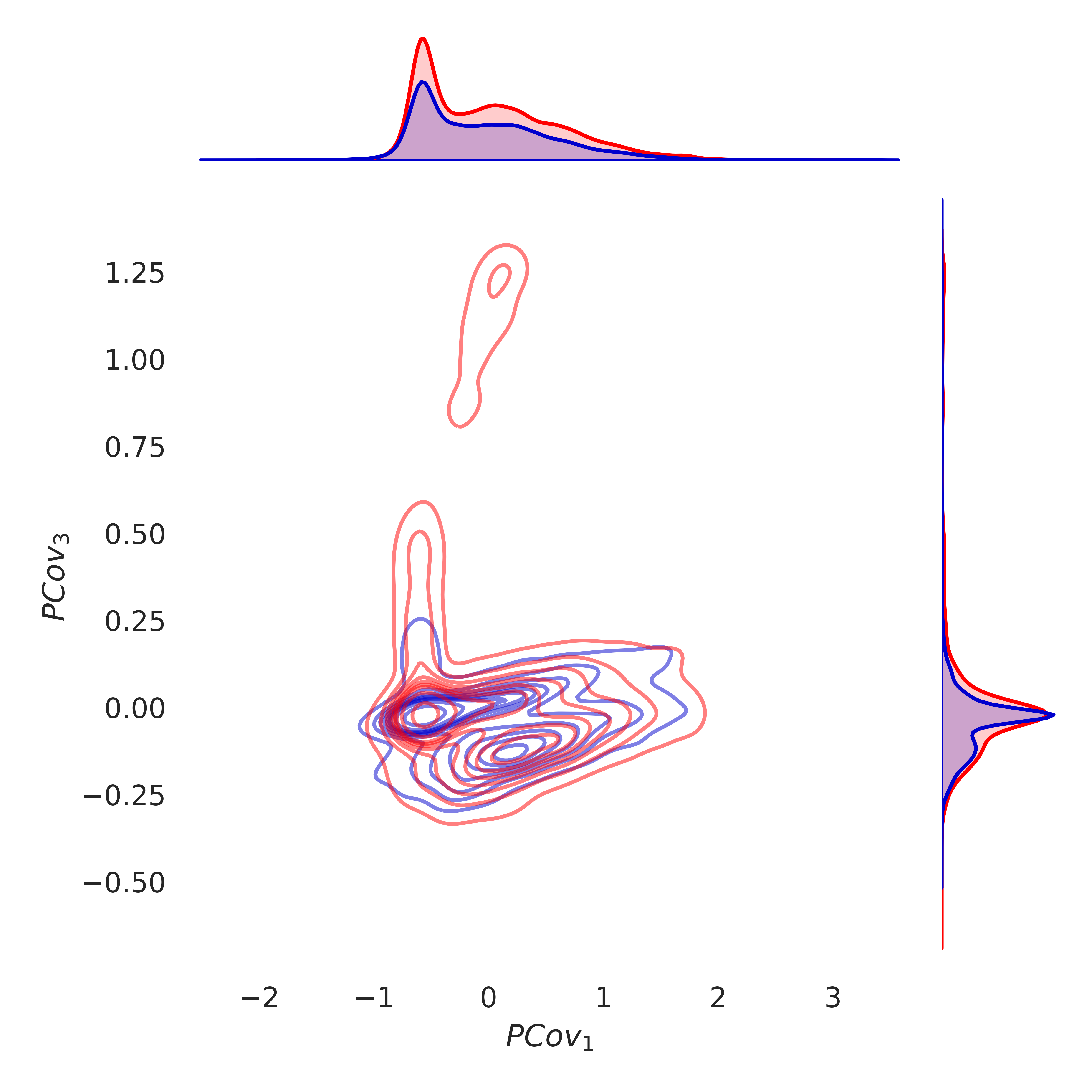}
\includegraphics[width=.45\columnwidth]{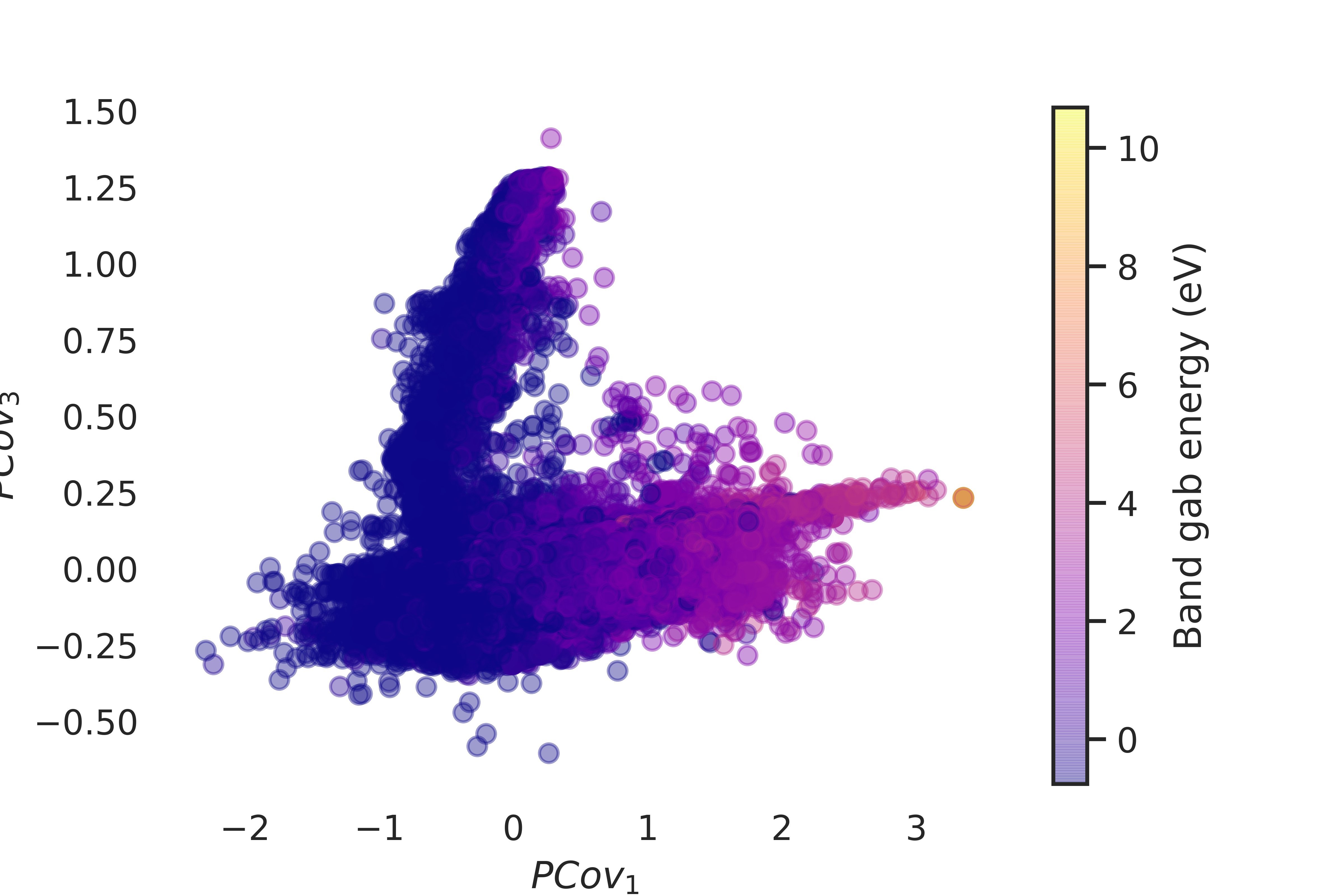}
\includegraphics[width=.45\columnwidth]{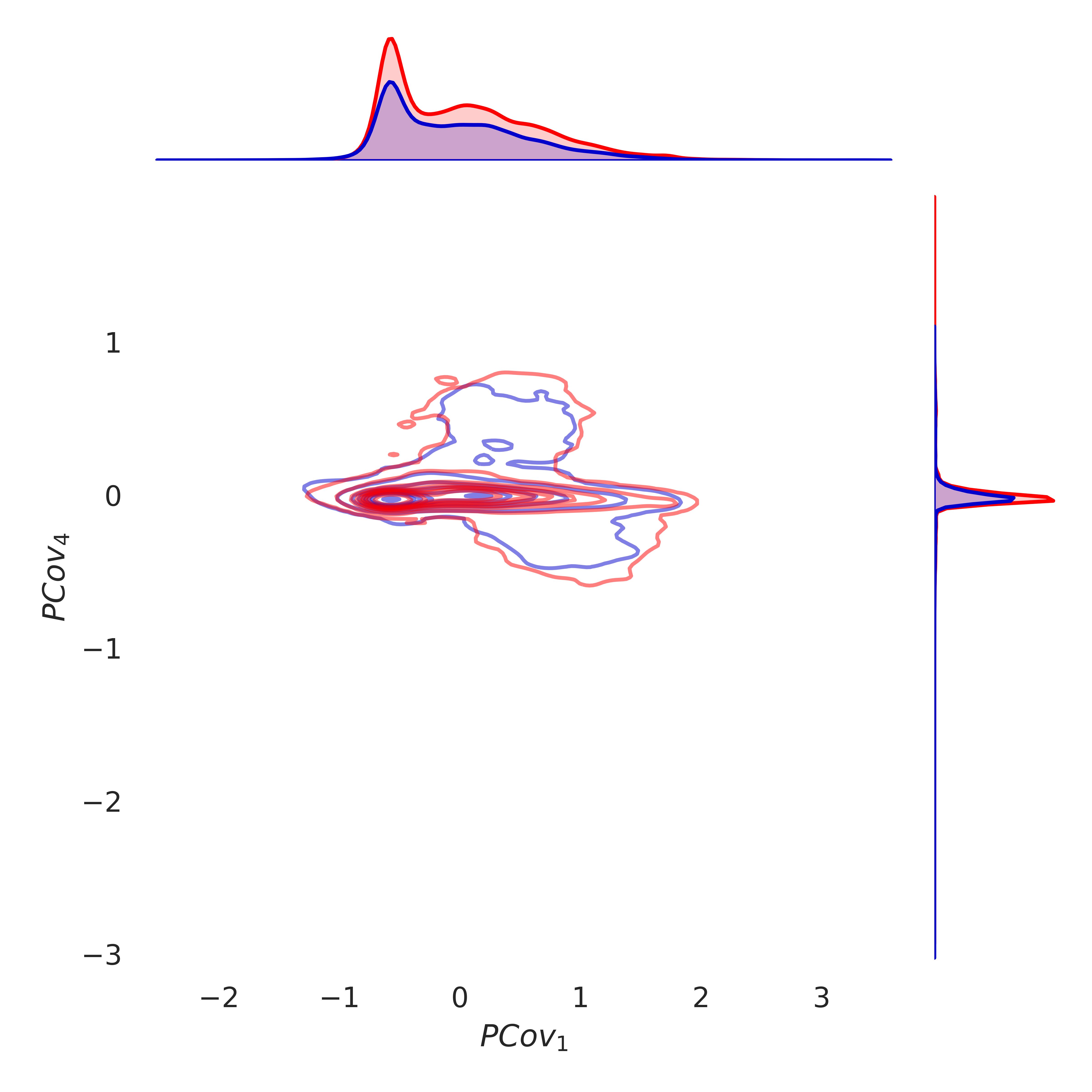}
\includegraphics[width=.45\columnwidth]{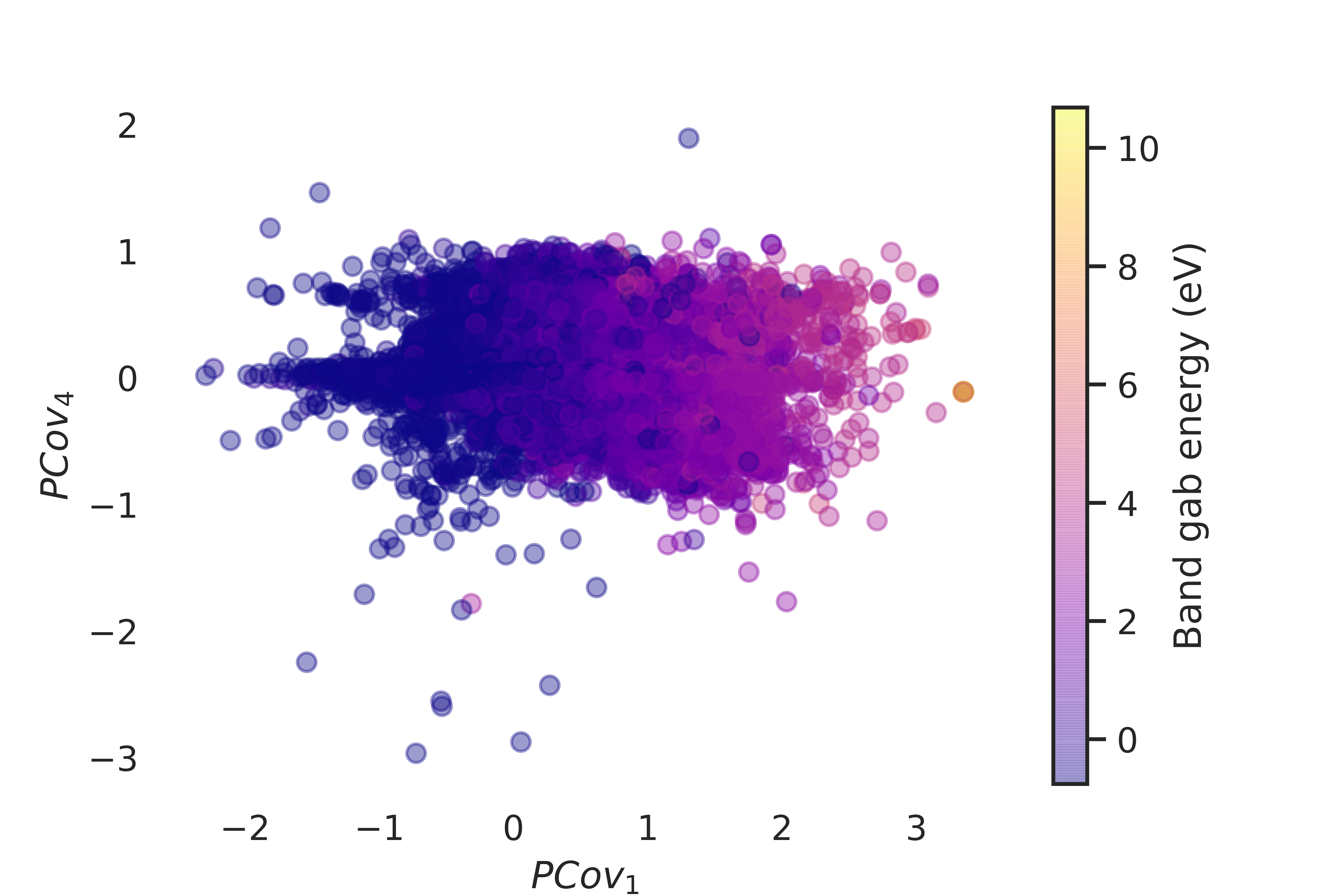}
\caption{
       PCovR representation with $\beta$ = 0.5 containing information on SOAP representation and regressed on the band gap energy. }
\label{fig:band_e}
\end{figure*}

The following section explores the PCovR energetic analysis performed on the MP dataset with the aim of classifying the structures into the two subgroups by performing a linear regression on local energetic descriptors only. Different covariates are plotted against the first principal covariate (on the $x$ axis each time) to explore the full database variance. Each image is reported in two different views: on the left, the compounds are coloured according to their energetic property, i.e. formation energy per atom (Figure \ref{fig:e_form}), energy above the convex hull (Figure \ref{fig:convex_hull}) and band gap energy (Figure \ref{fig:band_e}), while on the right the same data is coloured according to the subset it belongs to using a kernel density probability estimation (KDE) normalised to the whole set of data.  
The isolated areas containing only \textit{magic} structures mostly contain structures with Mg-O square bonds, ionic bonds with high bond energy and therefore lower formation energy per atom. They validate the SOAP representation's usefulness in separating between compounds' subgroups, but are not enough to draw insightful conclusions on the RoF.
% The top panel on the right of the figure is in agreement with the consideration relating Mg-O square bonds with low formation energy, being them characterized by a high $PF$.

% \begin{figure*}[htbp!] 
%   \centering
%   \includegraphics[width=1\columnwidth]{energies/form_e.png}
%   \caption{
%     PCovR representation with $\beta$ = 0.5 containing information on SOAP representation and regressed on the formation energy per atom. } 
%   \label{fig:e_form}
% \end{figure*}

% \begin{figure*}[htbp!] 
%   \centering
%   \includegraphics[width=1\columnwidth]{energies/convex_e.png}
%   \caption{
%     PCovR representation with $\beta$ = 0.5 containing information on SOAP representation and regressed on the energy above the convex hull. } 
%   \label{fig:convex_e}
% \end{figure*}

% \begin{figure*}[htbp!] 
%   \centering
%   \includegraphics[width=1\columnwidth]{energies/band_e.png}
%   \caption{
%     PCovR representation with $\beta$ = 0.5 containing information on SOAP representation and regressed on the band gap energy. } 
%   \label{fig:band_e}
% \end{figure*}
\newpage

\section*{V. Classification algorithms}

The statistics of the accuracy on the test set achieved by the different classification methods are reported in Table \ref{table:classif}. The Random Forest classifier \cite{breiman2001random} is not only the best-performing classifier, but has the smallest discrepancy between using species-invariant and species-tagged representations, which implies that the classification is primarily embedded in the local symmetries, despite the species information.

\begin{table}[htbp!]
\centering
\begin{tabular}{|m{1.5cm}|m{3.0cm}|m{3cm}|m{7.5cm}|}
\hline
Classifier & Test Set R$^2$ & Test Set R$^2$ &Classifier parameters\\
 & (Species-Invariant) & (Species-Tagged) &\\
\hline
\hline
Random Forest & 0.871 & 0.88 &\{\texttt{bootstrap}: True, \texttt{ccp alpha}: 0.0, \texttt{class weight}: None, \texttt{criterion}: gini, \texttt{max depth}: None, \texttt{max features}: sqrt, \texttt{max leaf nodes}: None, \texttt{max samples}: None, \texttt{min impurity decrease}: 0.0, \texttt{min samples leaf}: 1, \texttt{min samples split}: 2, \texttt{min weight fraction leaf}: 0.0, \texttt{n estimators}: 100, \texttt{n jobs}: 4, \texttt{oob score}: False, \texttt{random state}: 2, \texttt{verbose}: 2, \texttt{warm start}: False\}\\
\hline
MLP Classifier & 0.728 & 0.801 &\{\texttt{activation}: relu, \texttt{alpha}: 0.0001, \texttt{batch size}: auto, \texttt{beta 1}: 0.9, \texttt{beta 2}: 0.999, \texttt{early stopping}: False, \texttt{epsilon}: 1e-08, \texttt{hidden layer sizes}: (100,), \texttt{learning rate}: constant, \texttt{learning rate init}: 0.001, \texttt{max fun}: 15000, \texttt{max iter}: 200, \texttt{momentum}: 0.9, \texttt{n iter no change}: 10, \texttt{nesterovs momentum}: True, \texttt{power t}: 0.5, \texttt{random state}: 2, \texttt{shuffle}: True, \texttt{solver}: adam, \texttt{tol}: 0.0001, \texttt{validation fraction}: 0.1, \texttt{verbose}: 2, \texttt{warm start}: False\}\\
\hline
Decision Tree & 0.587 & 0.805 &\{\texttt{ccp alpha}: 0.0, \texttt{class weight}: None, \texttt{criterion}: gini, \texttt{max depth}: None, \texttt{max features}: 80, \texttt{max leaf nodes}: None, \texttt{min impurity decrease}: 0.0, \texttt{min samples leaf}: 1, \texttt{min samples split}: 2, \texttt{min weight fraction leaf}: 0.0, \texttt{random state}: 2, \texttt{splitter}: best\}\\
\hline
Linear SVM & 0.594 & 0.67 &\{\texttt{C}: 1.0, \texttt{class weight}: None, \texttt{dual}: True, \texttt{fit intercept}: True, \texttt{intercept scaling}: 1, \texttt{loss}: squared hinge, \texttt{max iter}: 1000, \texttt{multi class}: ovr, \texttt{penalty}: l2, \texttt{random state}: 2, \texttt{tol}: 0.0001, \texttt{verbose}: 2\}\\
\hline
Cross-Validated Logistic Regression & 0.628 & 0.677 &\{\texttt{Cs}: 10, \texttt{class weight}: None, \texttt{cv}: 2, \texttt{dual}: False, \texttt{fit intercept}: True, \texttt{intercept scaling}: 1.0, \texttt{l1 ratios}: None, \texttt{max iter}: 100, \texttt{multi class}: auto, \texttt{n jobs}: 4, \texttt{penalty}: l2, \texttt{random state}: 2, \texttt{refit}: True, \texttt{scoring}: None, \texttt{solver}: lbfgs, \texttt{tol}: 0.0001, \texttt{verbose}: 2\}\\
\hline
Stochastic Gradient Descent Classifier & 0.594 & 0.61 &\{\texttt{alpha}: 0.0001, \texttt{average}: False, \texttt{class weight}: None, \texttt{early stopping}: False, \texttt{epsilon}: 0.1, \texttt{eta0}: 0.0, \texttt{fit intercept}: True, \texttt{l1 ratio}: 0.15, \texttt{learning rate}: optimal, \texttt{loss}: hinge, \texttt{max iter}: 100, \texttt{n iter no change}: 5, \texttt{n jobs}: 4, \texttt{penalty}: l2, \texttt{power t}: 0.5, \texttt{random state}: 2, \texttt{shuffle}: True, \texttt{tol}: 0.001, \texttt{validation fraction}: 0.1, \texttt{verbose}: 2, \texttt{warm start}: False\}\\
% \hline
% % QDA & 0.593 & 0.606 &\{\texttt{priors}: None, \texttt{reg param}: 0.0, \texttt{store covariance}: False, \texttt{tol}: 0.0001\}\\
% % \hline
% Gaussian Naive Bayes & 0.59 & 0.63 &\{\texttt{priors}: None, \texttt{var smoothing}: 1e-09\}\\
\hline\end{tabular}
\caption{Accuracy on test set achieved by different classifiers.}  
\label{table:classif}
\end{table}

\newpage

\newpage

% \bibliographystyle{naturemag}
% \bibliography{bib.bib}

% \printbibliography
\bibliography{bib.bib}

% \bibliography{si}